# Network visualisations related to special functions based on the Scopus data since 1940

Rushan Ziatdinov[a]

[a]Department of Industrial Engineering, College of Engineering, Keimyung University, 704-701 Daegu, Republic of Korea

**ABSTRACT**
Special functions are essential in theoretical and applied mathematics and have various applications in the applied sciences. Mathematicians have studied them for centuries, but there is still no bibliometric analysis that summarises the datasets of publications showing different network visualisations, such as co-author and keyword visualisations, basic keyword statistics and other data analyses. This work appears to be the first attempt to fill this gap by presenting different network visualisations based on 4025 documents with the keyword "special function" in their title, abstract or keywords belonging to the field of mathematics in the Scopus database. We also show that special functions are rarely used in geometric modelling, a mathematical foundation for CAD, industrial design, architecture, and other applied fields, and we discuss how different visualisations for special functions can be generated in the Julia programming language. The generated image and video visualisations can be helpful to academics to see the impact of different authors, to define their new research topics, to see connections or lack thereof between various topics, to find the most popular special functions, or for teaching purposes to show the importance of special functions and links to other topics in modern pure and applied mathematics and other sciences.

**KEYWORDS**
Special function; visualisation; network; visual analytics; keyword analysis; co-authorship; VOSviewer; Scopus database

## 1. Introduction

Special functions [1–3] play an essential role in theoretical and applied mathematics and have various applications in applied sciences, such as different engineering fields: information theory, image processing, aesthetic shape modelling, computer simulation, etc.

Mathematicians have studied them for several centuries, but there is still no bibliometric analysis that summarises the datasets of related publications showing different network visualisations, such as co-author and keyword visualisations, basic keyword statistics and other data analyses. This work seems to be the first attempt to fill this gap by presenting different network visualisations based on 4025 Scopus-indexed docu-

CONTACT Rushan Ziatdinov. Email: ziatdinov@kmu.ac.kr, ziatdinov.rushan@gmail.com, URL: https://www.ziatdinov-lab.com/

ments from the field of mathematics that have the keyword "special function" in their title, abstract or keywords.

The images generated by the VOSviewer software and the ScienceScape online tool, as well as the video visualisations, can help academics define their new research topics, see connections or their absence between different topics, find the most popular special functions in the literature, or, for teaching or presentation purposes, show the importance of special functions and their links (connections) to other topics in modern pure and applied mathematics and other sciences.

## 2. Methods

The Scopus search query TITLE-ABS-KEY("special function") AND (LIMIT-TO(SUBJAREA, "MATH")) in title, abstract, and keywords returned a list of 4025 documents exported from Scopus as a RIS file. For the data analysis, we used the open-source software VOS (https://www.vosviewer.com/), commonly used to construct and visualise bibliometric networks. In addition, we used ScienceScape (https://medialab.github.io/sciencescape/), an online tool that can generate networks and visualise data.

## 3. Results

### *3.1. General analysis of a graph*

Based on the information from VOSviewer, the network analysis reveals a substantial dataset with 2,939 nodes, 41,058 links and a total link strength of 56,259, indicating an extensive and interconnected network of research topics or publications. The network appears moderately dense, with an average of approximately 14 links per node, suggesting a significant degree of interconnectedness between the represented topics or publications. Furthermore, the total link strength (56,259) exceeds the total number of links (41,058). This means many links between nodes are reinforced, possibly indicating strong relationships or co-occurrences between specific topics or publications.

### *3.2. Most popular special functions*

According to our analysis, the most frequently mentioned special functions, sorted from most to least mentioned, are the Bessel function, hypergeometric function, gamma function, Mittag-Leffler function, Beta function, H-function, Riemann zeta function, Laguerre polynomials, Hermite polynomials, confluent hypergeometric function, Gauss hypergeometric function, Wright function, generalised hypergeometric function, Jacobi polynomials, incomplete gamma function, Appell function, error function, Humbert function, Legendre function, Bellman function, Gegenbauer polynomials, Legendre polynomials, Airy function, Polygamma function, Polylogarithm, $q$-special functions, associated Legendre functions, Chebyshev polynomials, Hurwitz zeta function, I-function, modified Bessel functions, parabolic cylinder functions, Whittaker functions, Bernoulli polynomials, Horn functions, Digamma function, exponential integral, Fox-Wright function, Struve function, Whittaker function.

The keyword evolution can be found online at `https://youtu.be/ht90Ou2EO6s`.

### 3.3. Most popular sources

According to Scopus, the most popular sources are Integral Transforms and Special Functions (168 papers), Journal of Mathematical Physics (103 papers), Journal of Mathematical Analysis and Applications (95 papers), Journal of Computational and Applied Mathematics (92 papers), Mathematics (73 papers), Lecture Notes in Computer Science (including subseries Lecture Notes in Artificial Intelligence and Lecture Notes in Bioinformatics) (67 papers), Applied Mathematics and Computation (66 papers), Proceedings of SPIE - The International Society for Optical Engineering (65 papers), Symmetry (62 papers), Mathematical Methods in the Applied Sciences (47 papers), Journal of Physics A: Mathematical and Theoretical (44 papers), Journal of Physics A: Mathematical and General (38 papers), Proceedings of the American Mathematical Society (36 papers), Ramanujan Journal (35 papers), Mathematics of Computation (31 papers), Computers and Mathematics with Applications (29 papers), AIMS Mathematics (26 papers), Axioms (25 papers), Fractal and Fractional (25 papers), Advances in Difference Equations (24 papers), Fractional Calculus and Applied Analysis (24 papers), Springer Proceedings in Mathematics and Statistics (23 papers), Symmetry, Integrability and Geometry: Methods and Applications (SIGMA) (23 papers), Journal of Inequalities and Applications (22 papers), Lobachevskii Journal of Mathematics (22 papers), Fractals (20 papers).

### 3.4. A-K-J Sankey graph

An A-K-J Sankey diagram is a special type of Sankey diagram used to visualise bibliometric data, particularly the relationships between authors (A), keywords (K) and journals (J) in the scientific literature. Figure 1 shows the flow and connections between these three key elements of scientific publications.

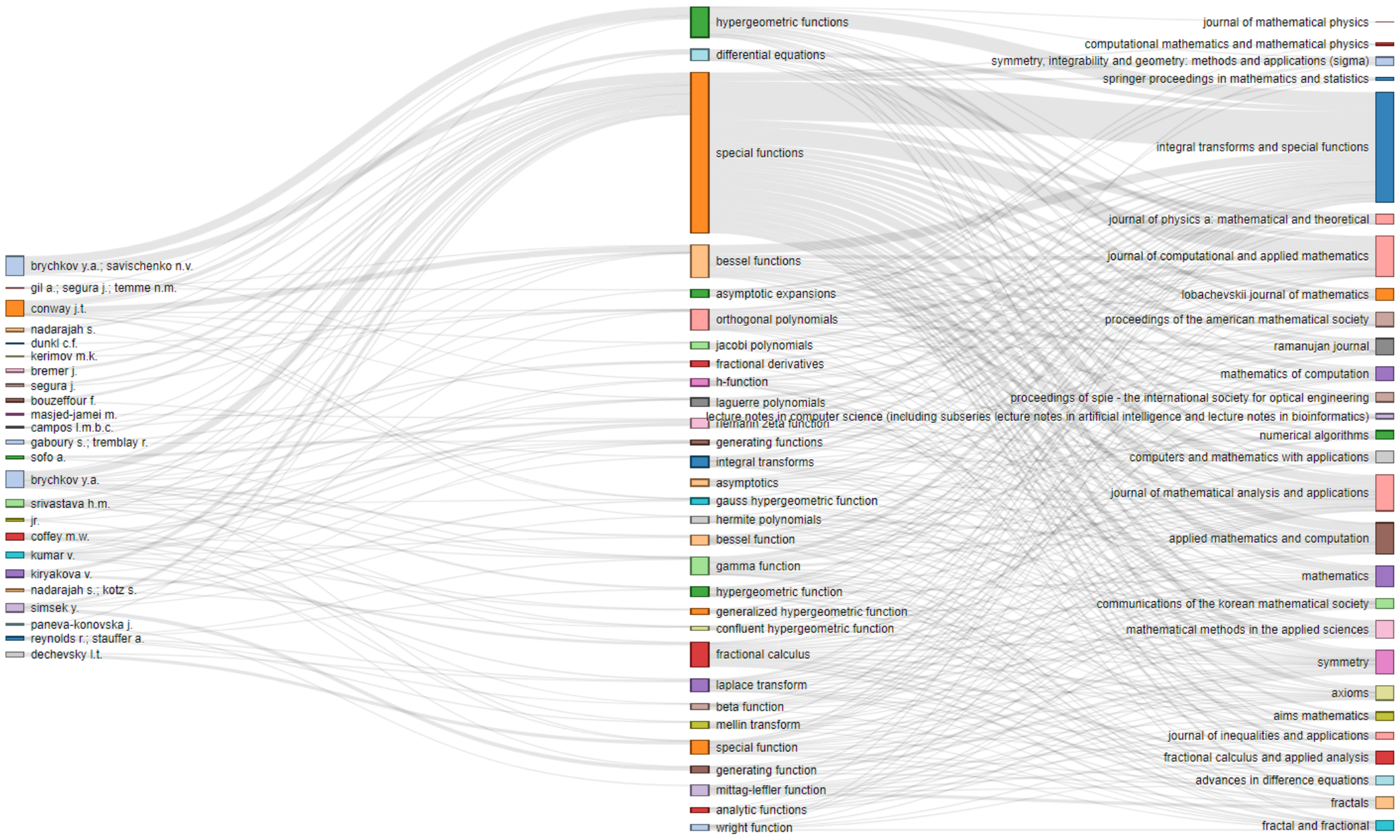

Figure 1.: Most active authors, keywords and journals.

### *3.5. Network visualisations*

#### *3.5.1. Network analysis*

We processed 2939 keywords with a minimum number of occurrences of 2, and the results are shown in this section. More detailed videos of the keyword and co-authorship networks with zoom can be found at `https://youtu.be/tgvRNqf4x8Q` and `https://youtu.be/Y_JXPO_KXx8`.

#### *3.5.2. Co-authorship analysis*

Figures 2 – 4 show a co-authorship network consisting of 1218 people with a minimum number of documents equal to two. The network consists of 391 clusters of authors.

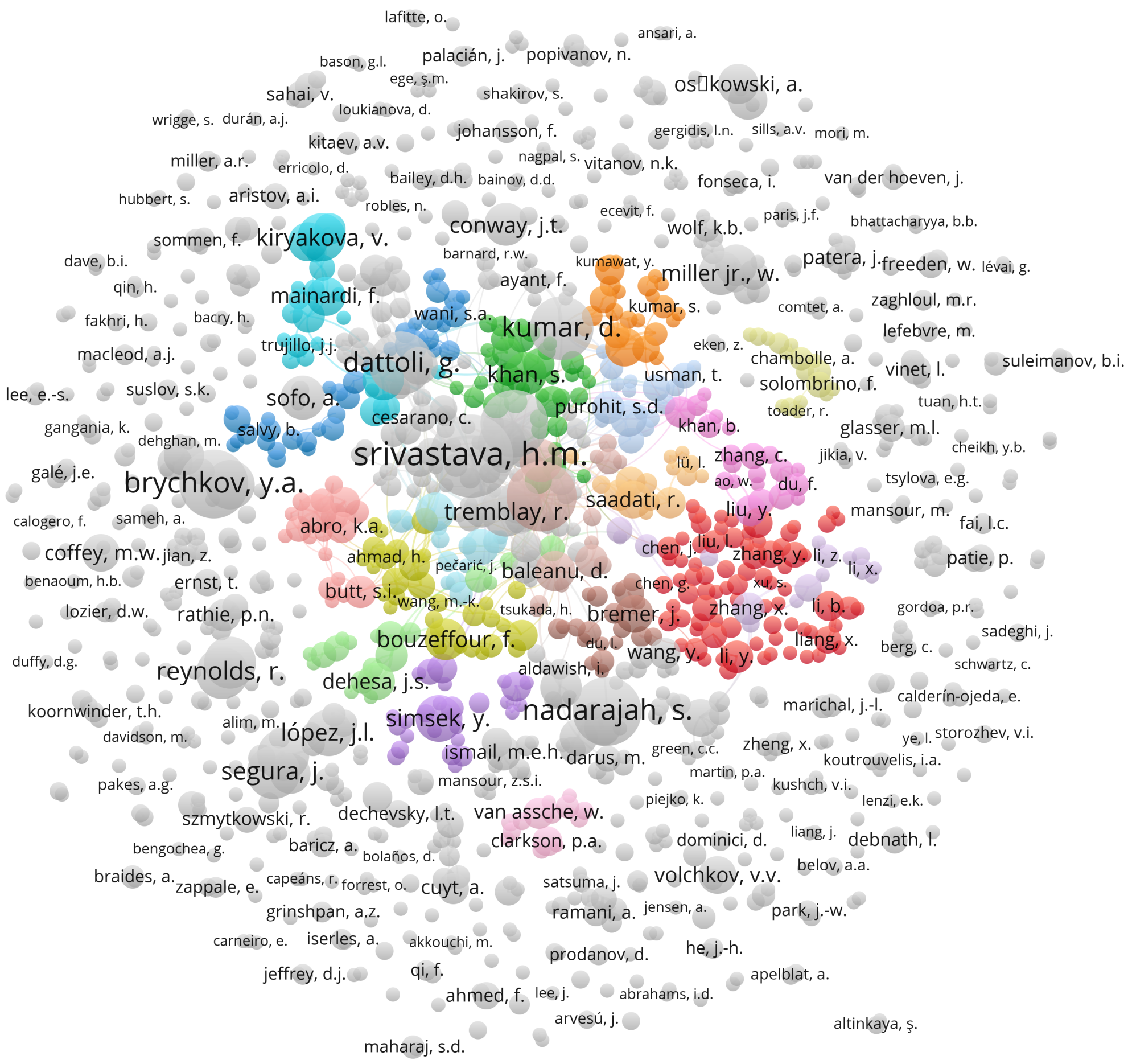

Figure 2.: Clusters of co-authors.

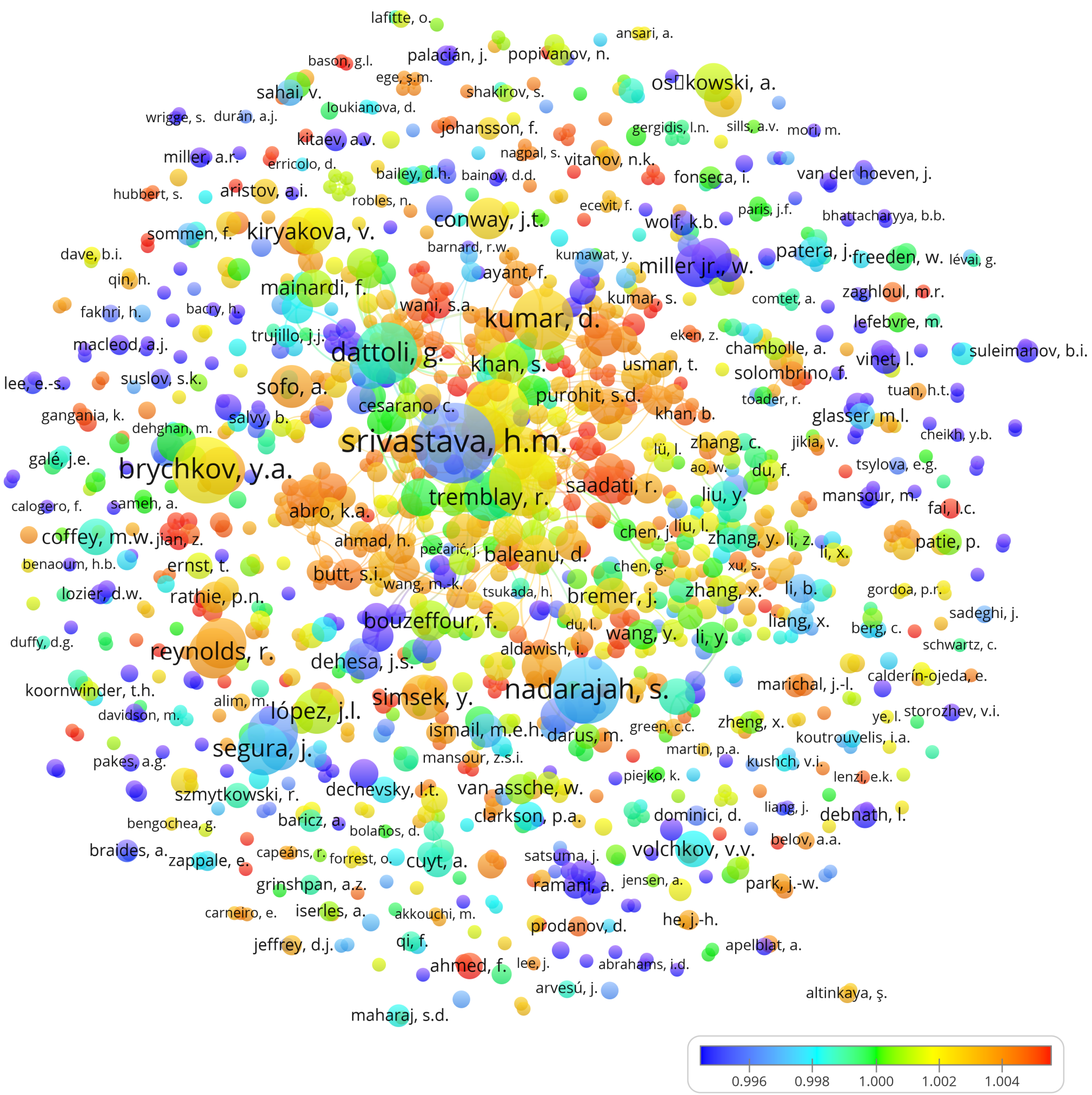

Figure 3.: Co-authorship visualisation by the number of documents. Cool colours indicate older documents, while warm colours indicate more recent work.

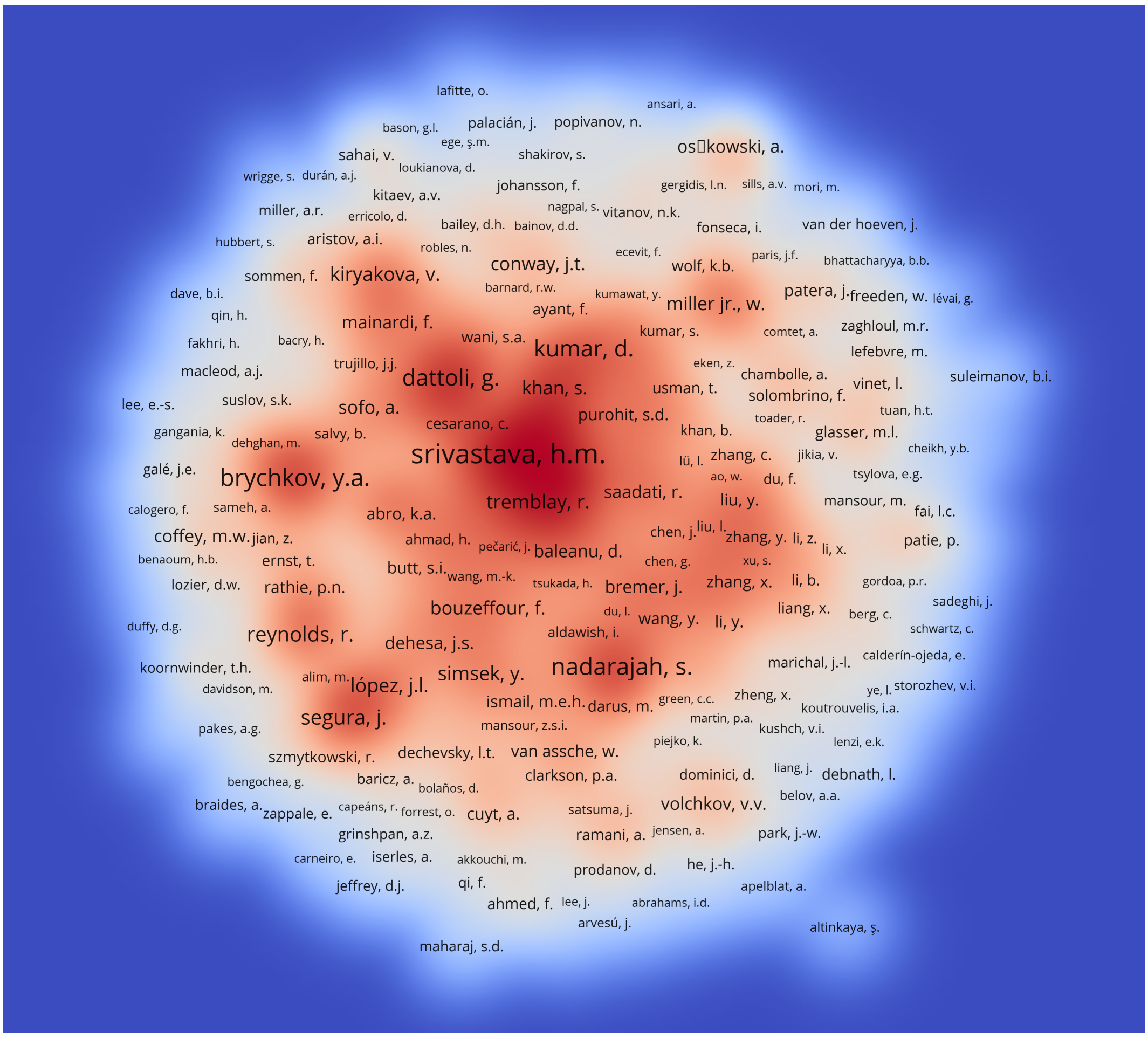

Figure 4.: Density visualisation of 1218 authors.

The following visualisations show the networks of the top ten authors. Some authors collaborate with many other researchers, while others work more individually and produce no less output regarding the number of documents.

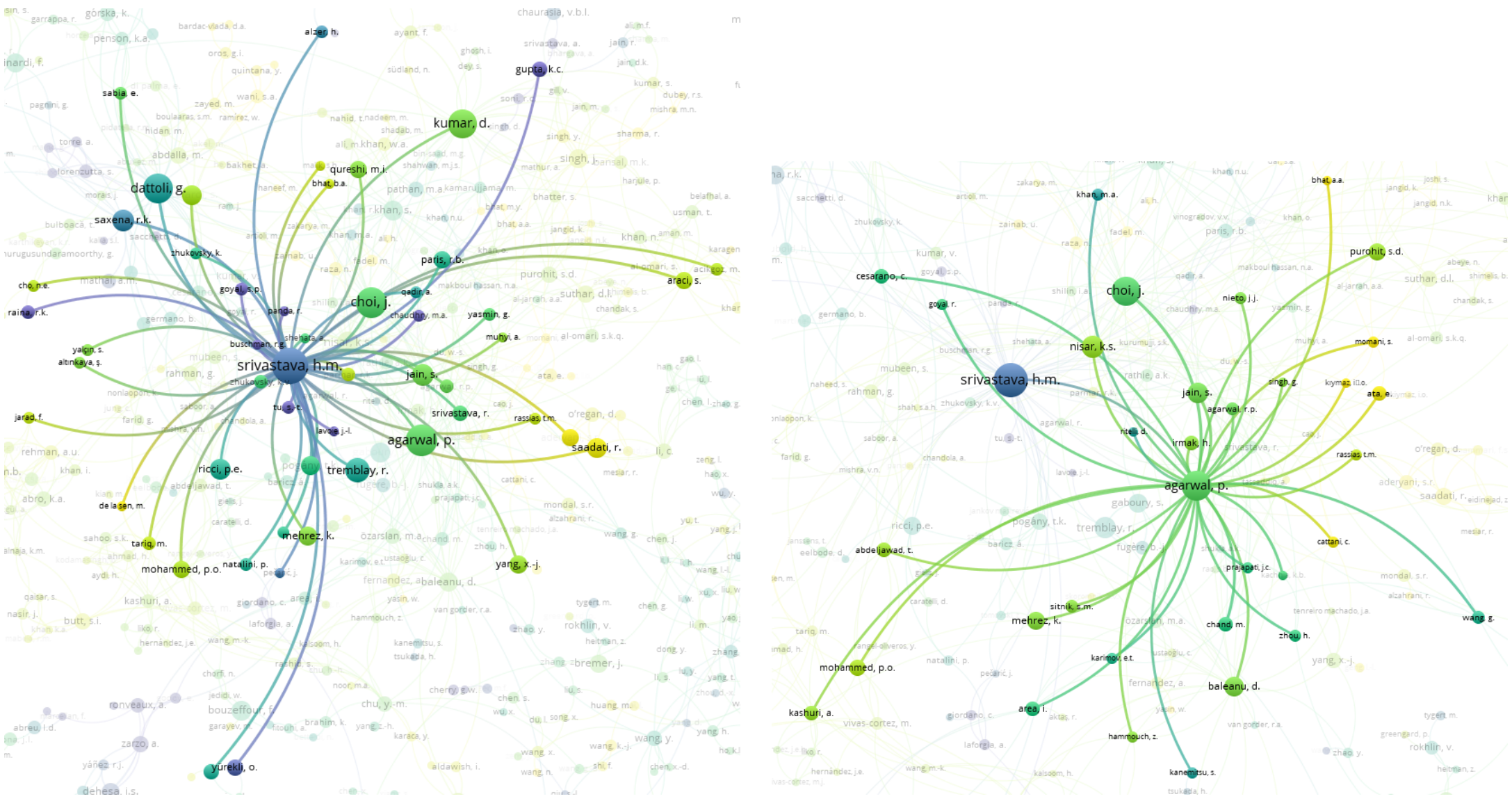

((a)) Co-authorship network of H.M. Srivastava.
((b)) Co-authorship network of P. Agarwal.

Figure 5.: Co-authorship networks. The lines show the links between different authors, and the size of the nodes is related to the number of documents containing the keyword ”special function” in the title, abstract or keywords.

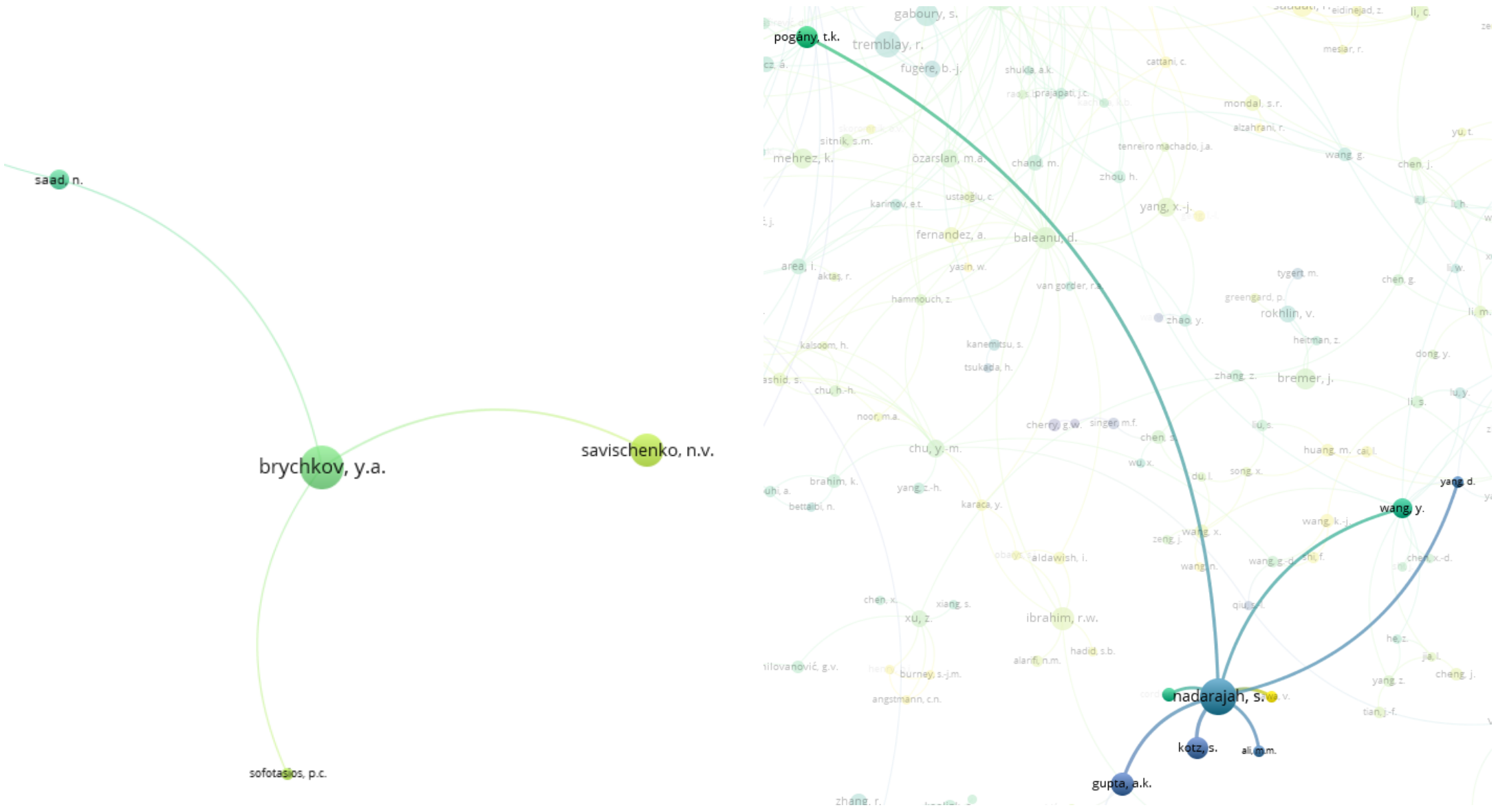

((a)) Co-authorship network of Y.A. Brychkov.
((b)) Co-authorship network of S. Nadarajah.

Figure 6.: Co-authorship networks. The lines show the links between different authors, and the size of the nodes is related to the number of documents containing the keyword ”special function” in the title, abstract or keywords.

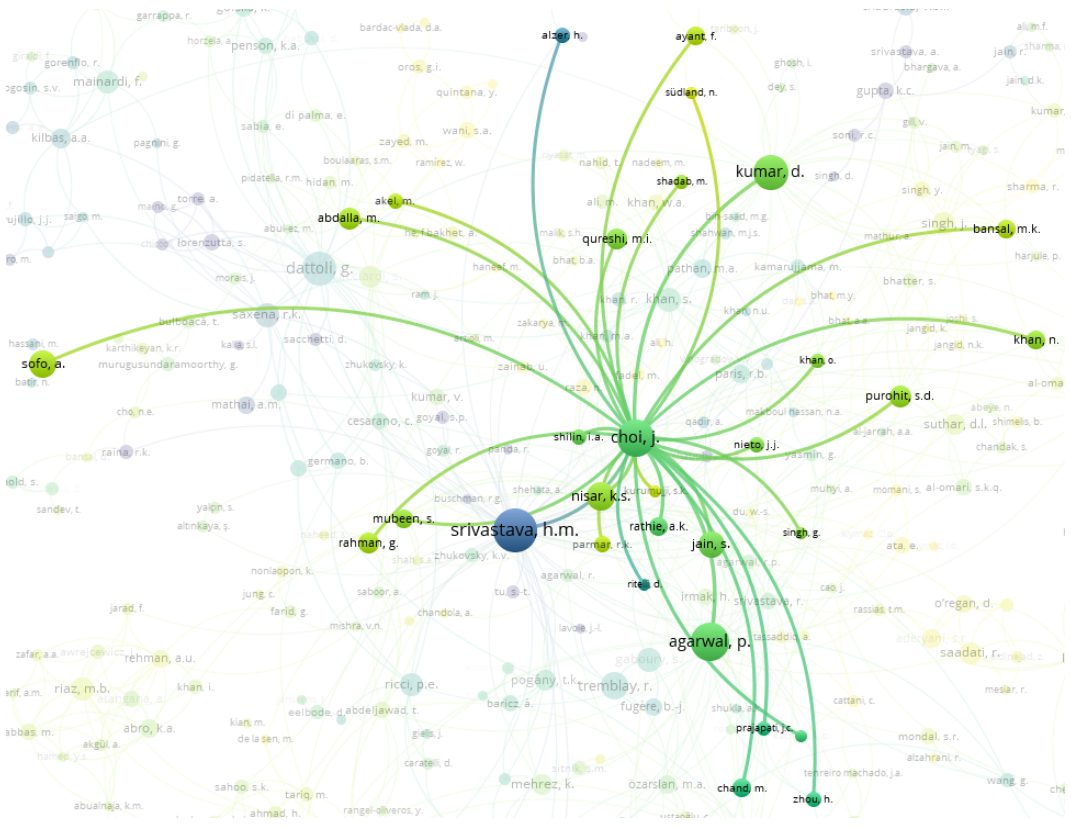

((a)) Co-authorship network of J. Choi.

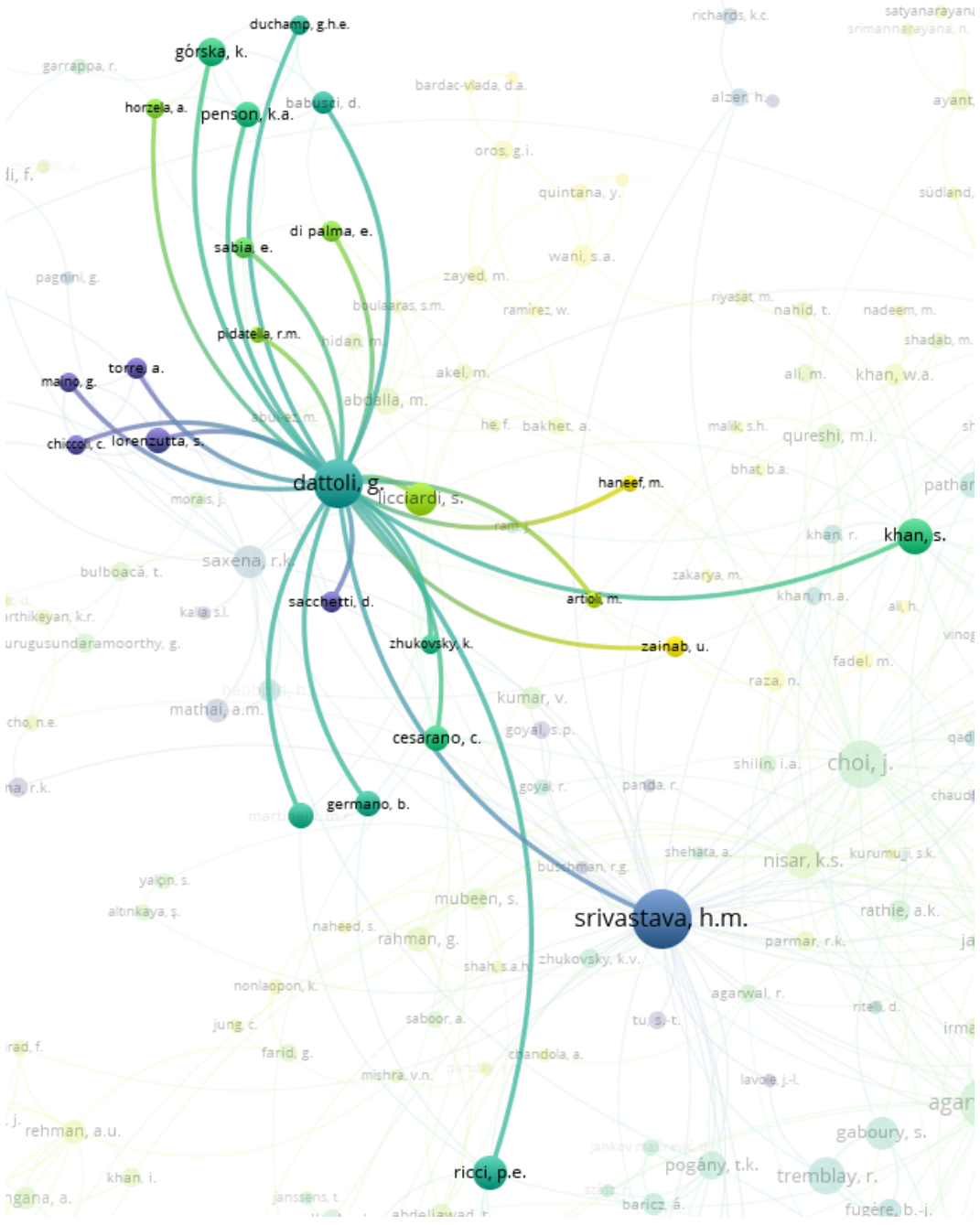

((b)) Co-authorship network of G. Dattoli.

Figure 7.: Co-authorship networks. The lines show the links between different authors, and the size of the nodes is related to the number of documents containing the keyword "special function" in the title, abstract or keywords.

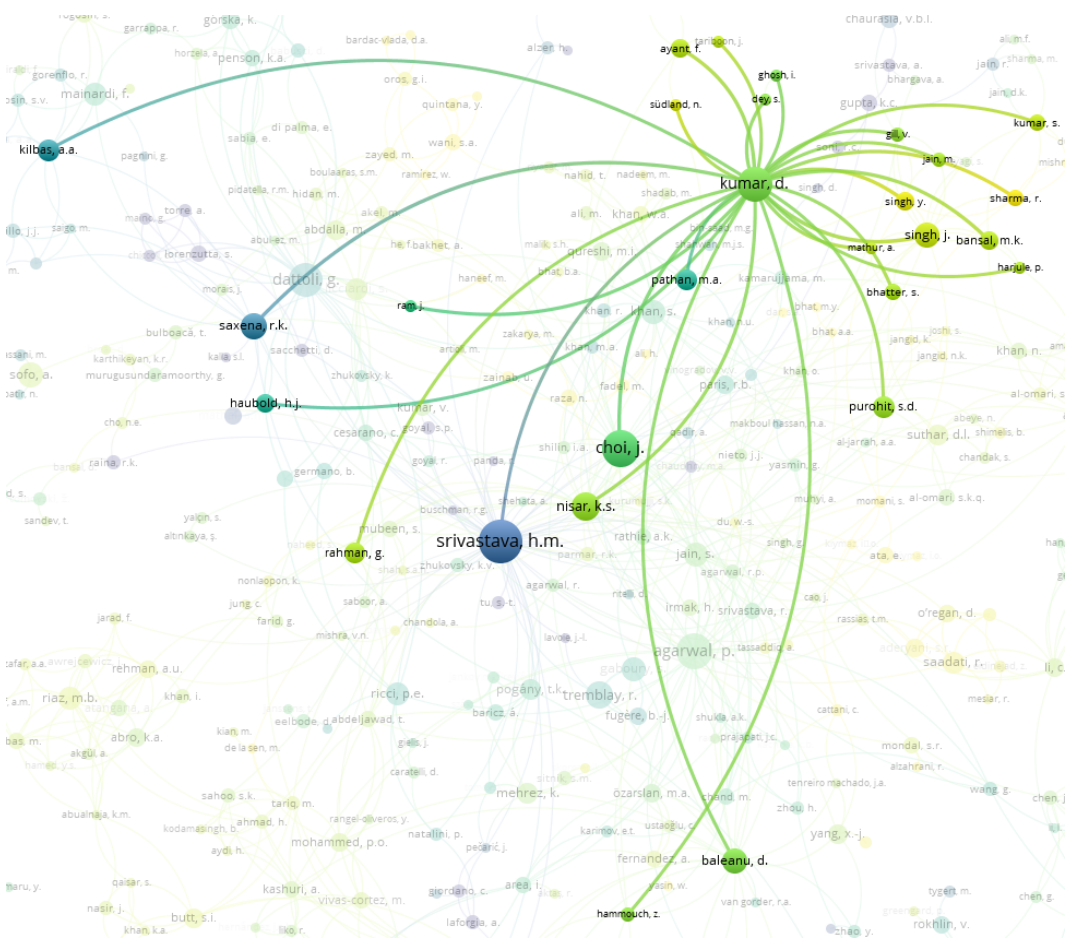

((a)) Co-authorship network of D. Kumar.

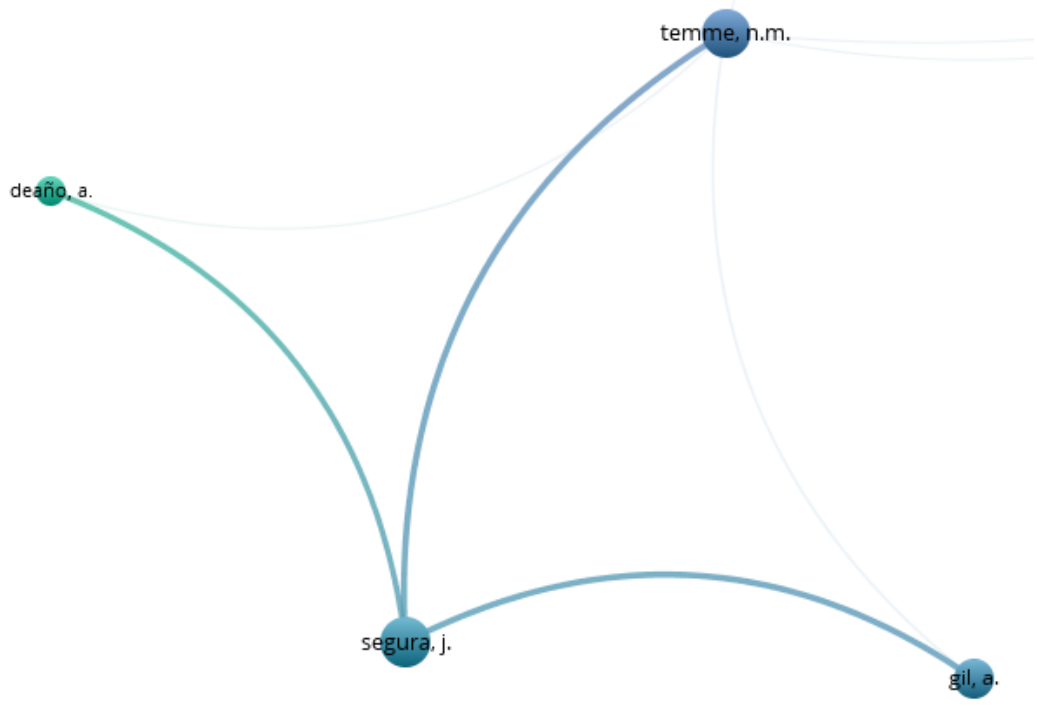

((b)) Co-authorship network of J. Segura.

Figure 8.: Co-authorship networks. The lines show the links between different authors, and the size of the nodes is related to the number of documents containing the keyword "special function" in the title, abstract or keywords.

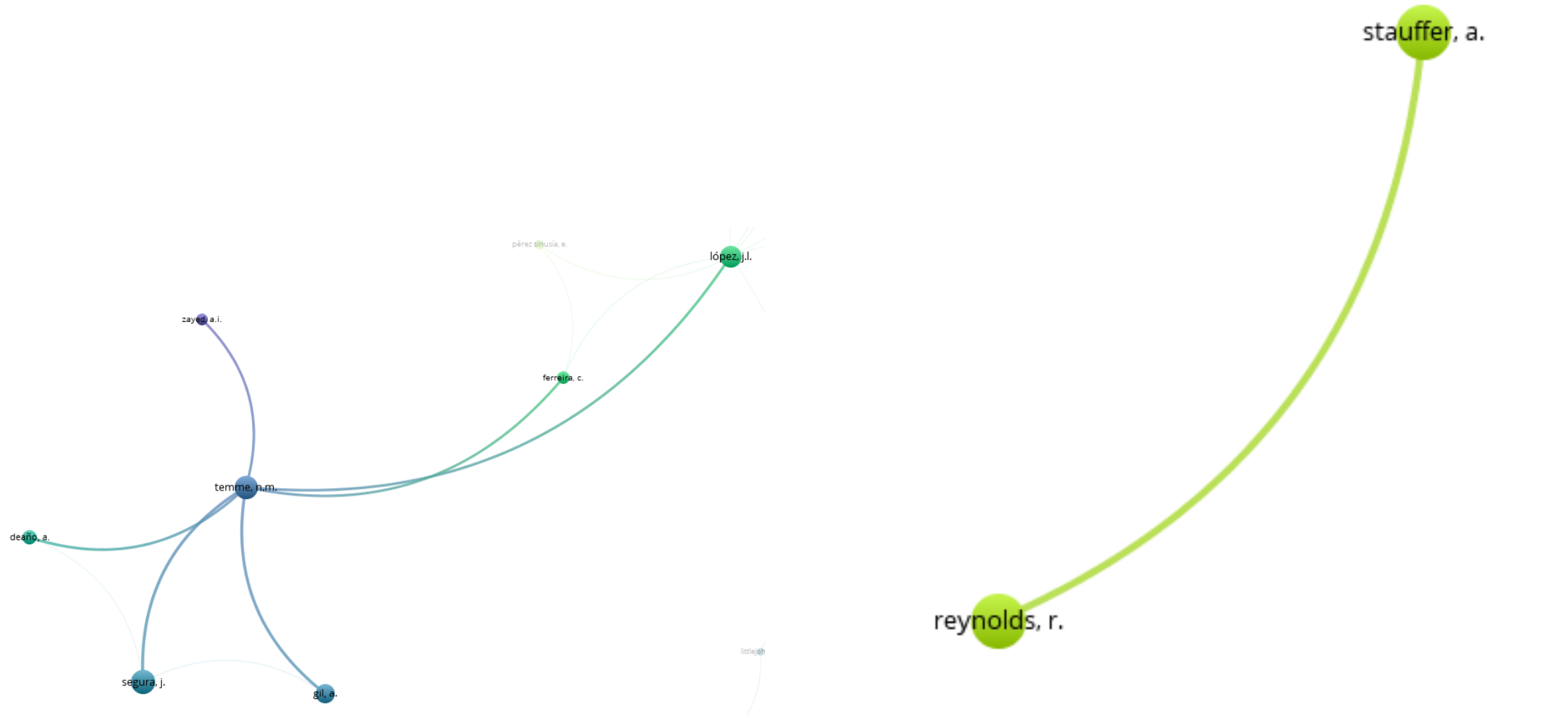

((a)) Co-authorship network of N.M. Temme.

((b)) Co-authorship network of R. Reynolds and A. Stauffer.

Figure 9.: Co-authorship networks. The lines show the links between different authors, and the size of the nodes is related to the number of documents containing the keyword ”special function” in the title, abstract or keywords.

### *3.5.3. Keyword analysis*

Figures 10–21 show visualisations of the networks for specific popular keywords by number of occurrences.

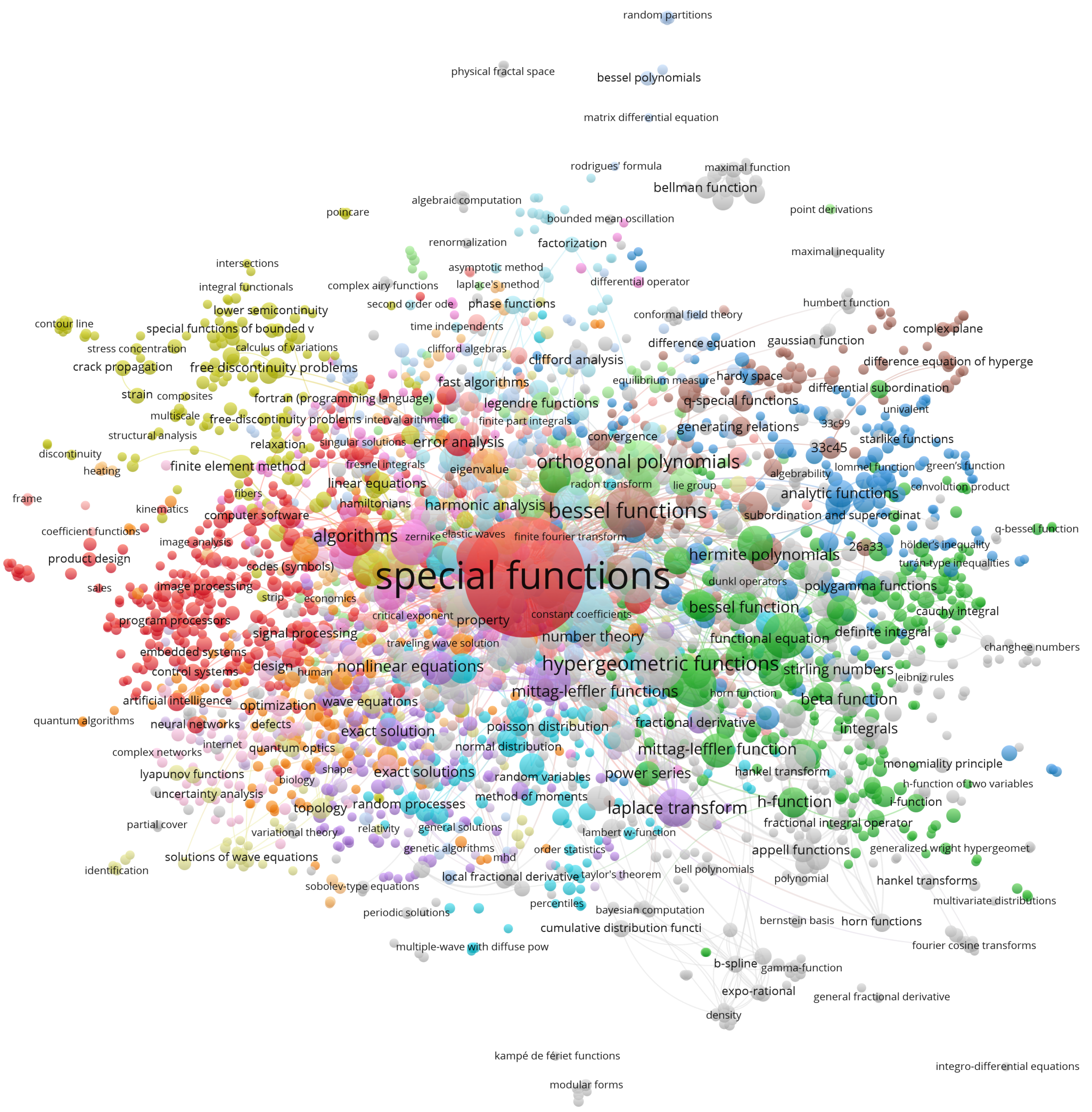

Figure 10.: A network visualisation of keywords and links by number of keyword occurrences. Here, different colours indicate 46 different clusters of keywords.

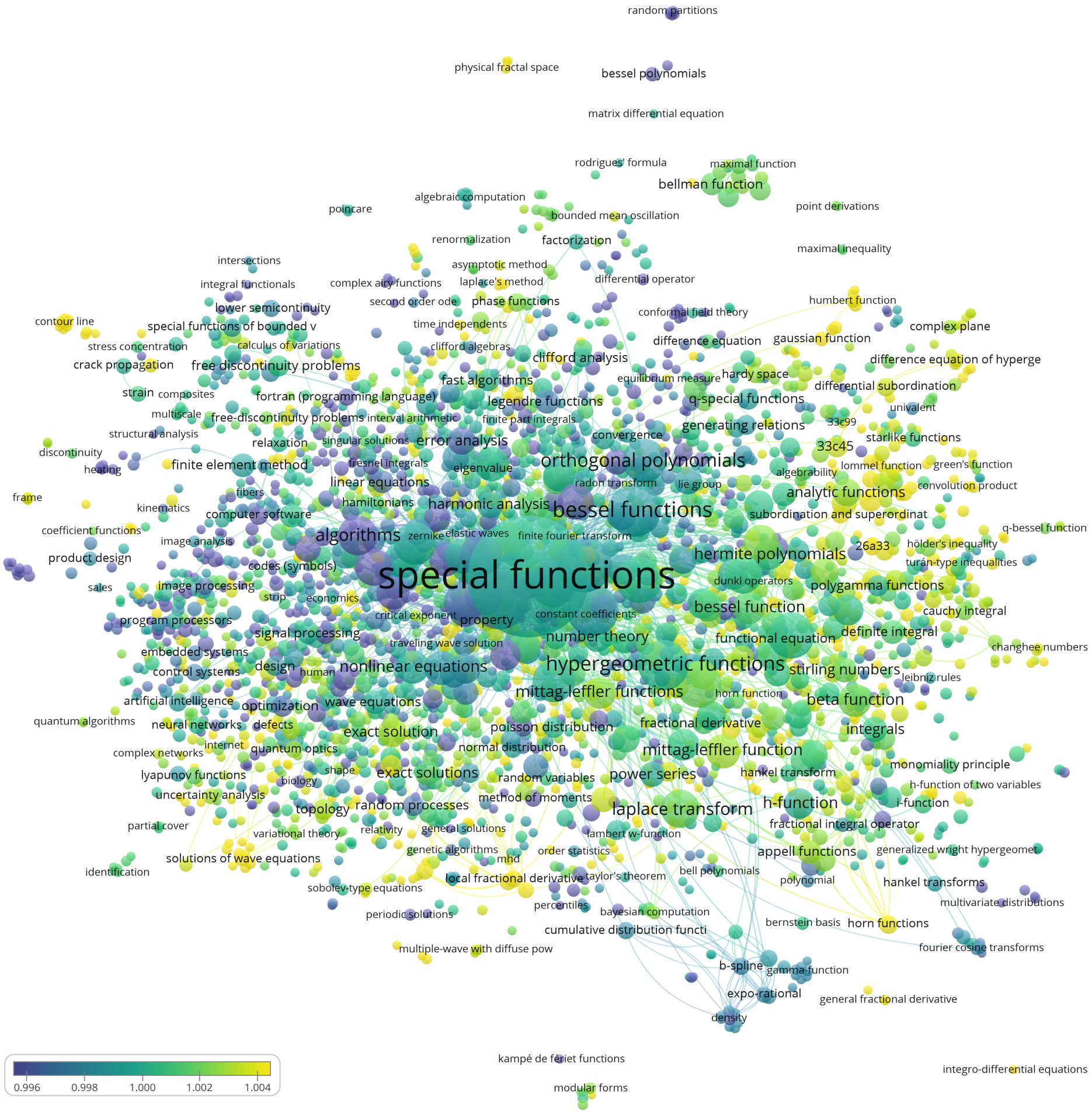

Figure 11.: An overlay visualisation of keywords and links by number of keyword occurrences. Here, the colour map is associated with normalised years from 1940 to 2025, i.e., blue indicates older works, and yellow indicates more recent works.

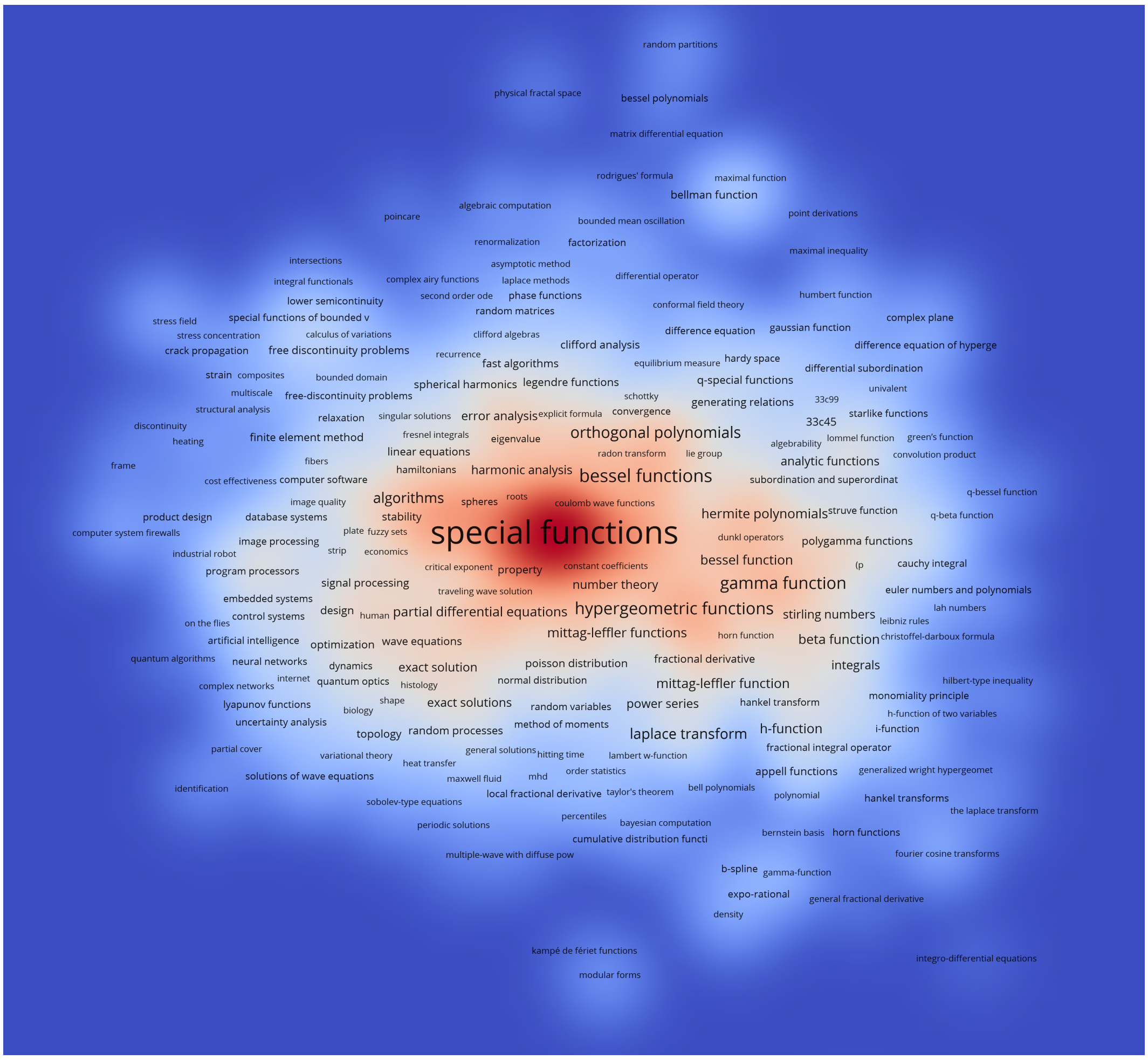

Figure 12.: A density visualisation of keywords by number of keyword occurrences.

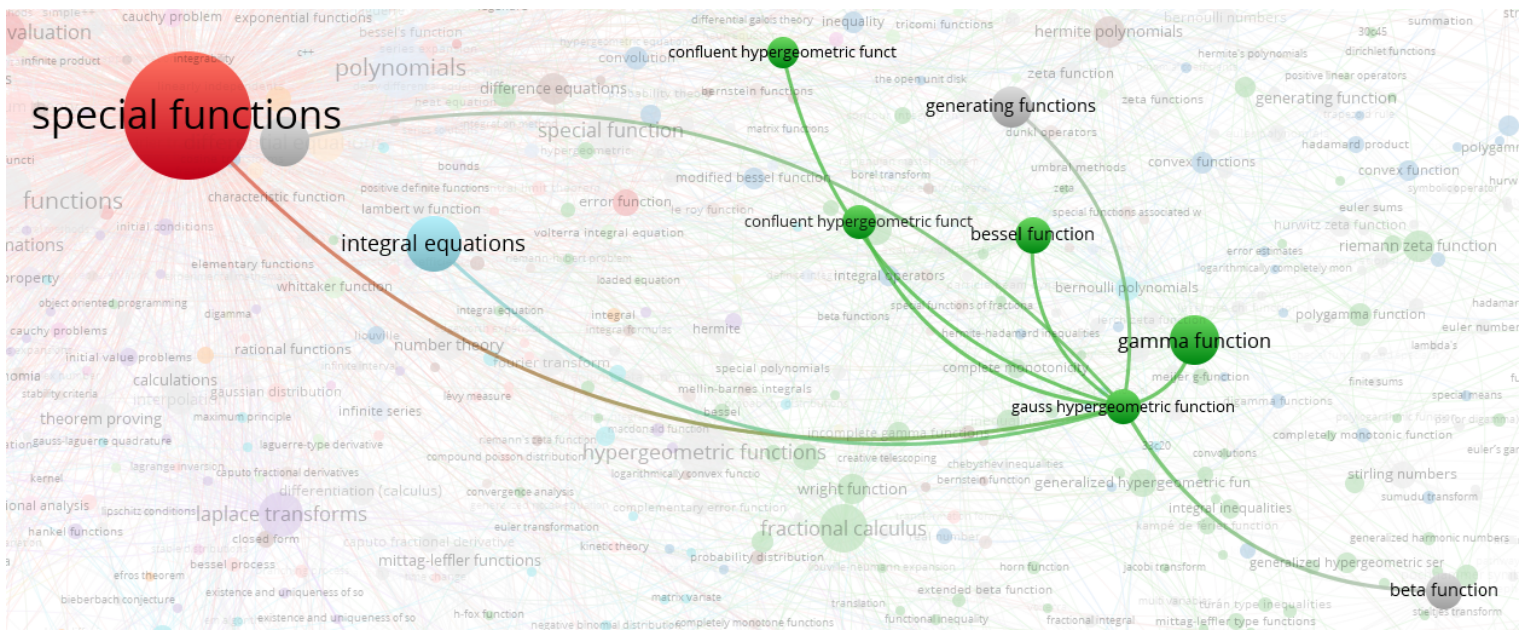

Figure 13.: The links of "Gauss hypergeometric function" in the studied keyword network.

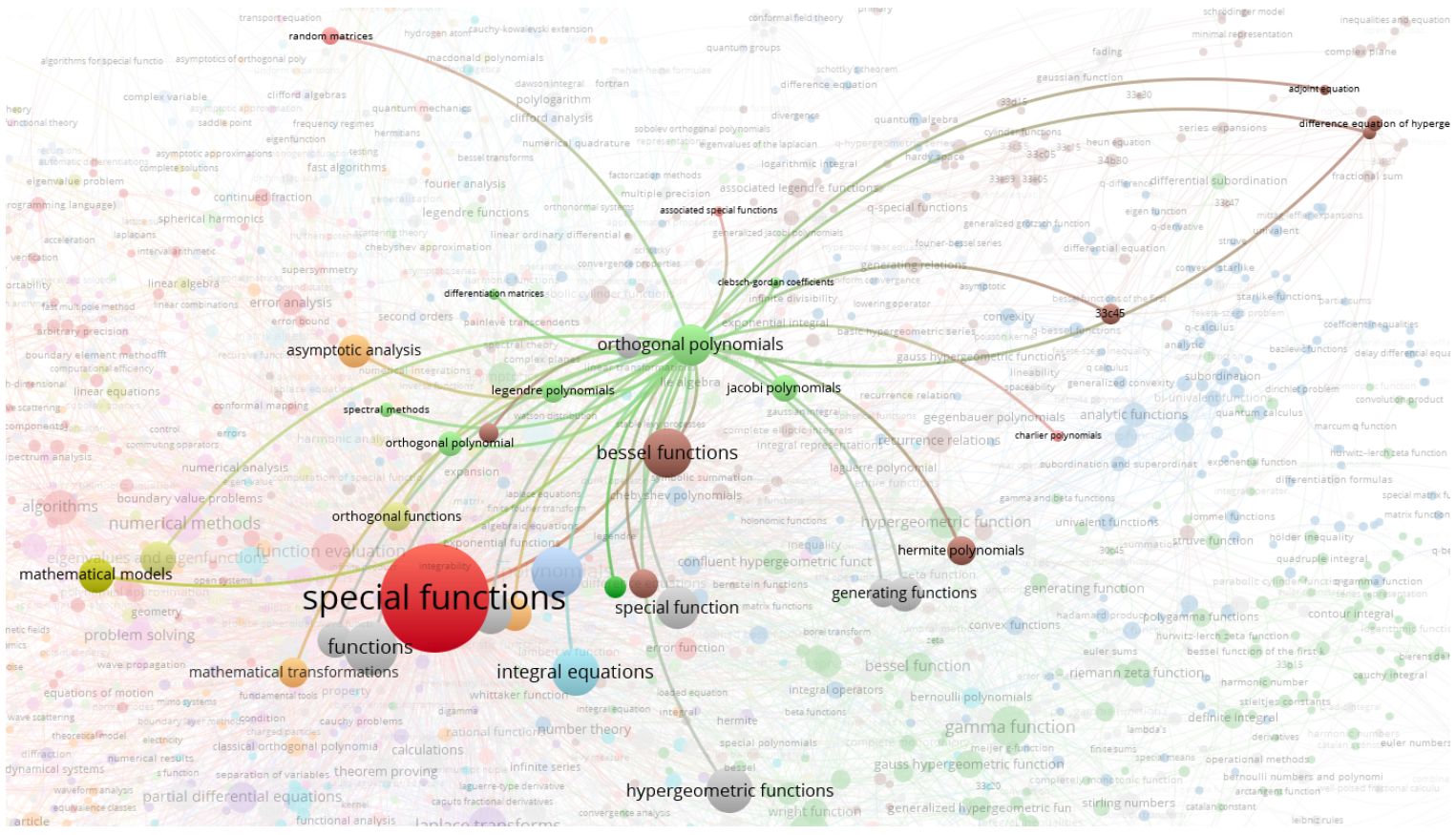

Figure 14.: The links of "orthogonal polynomials" in the studied keyword network.

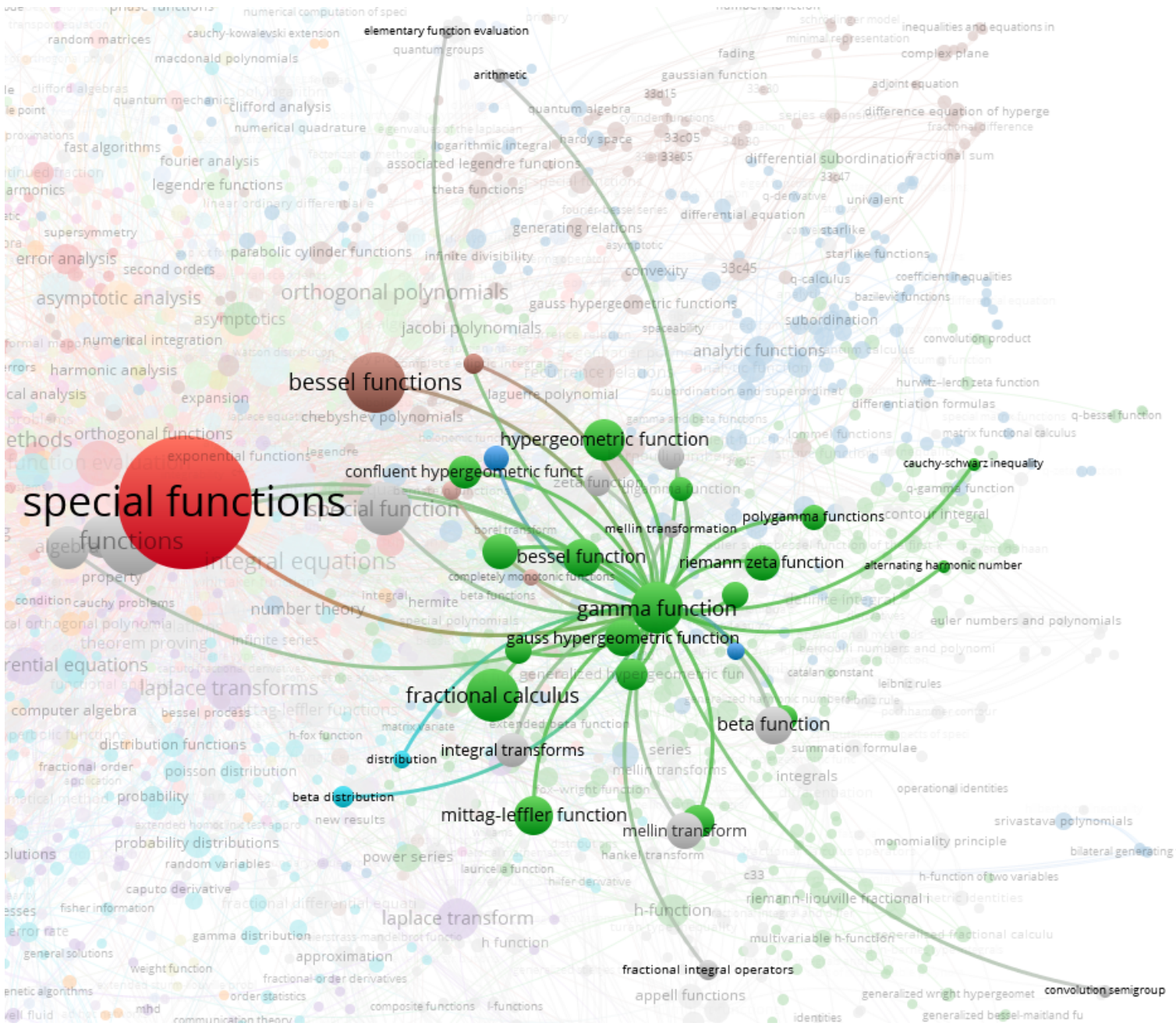

Figure 15.: The links of "gamma function" in the studied keyword network.

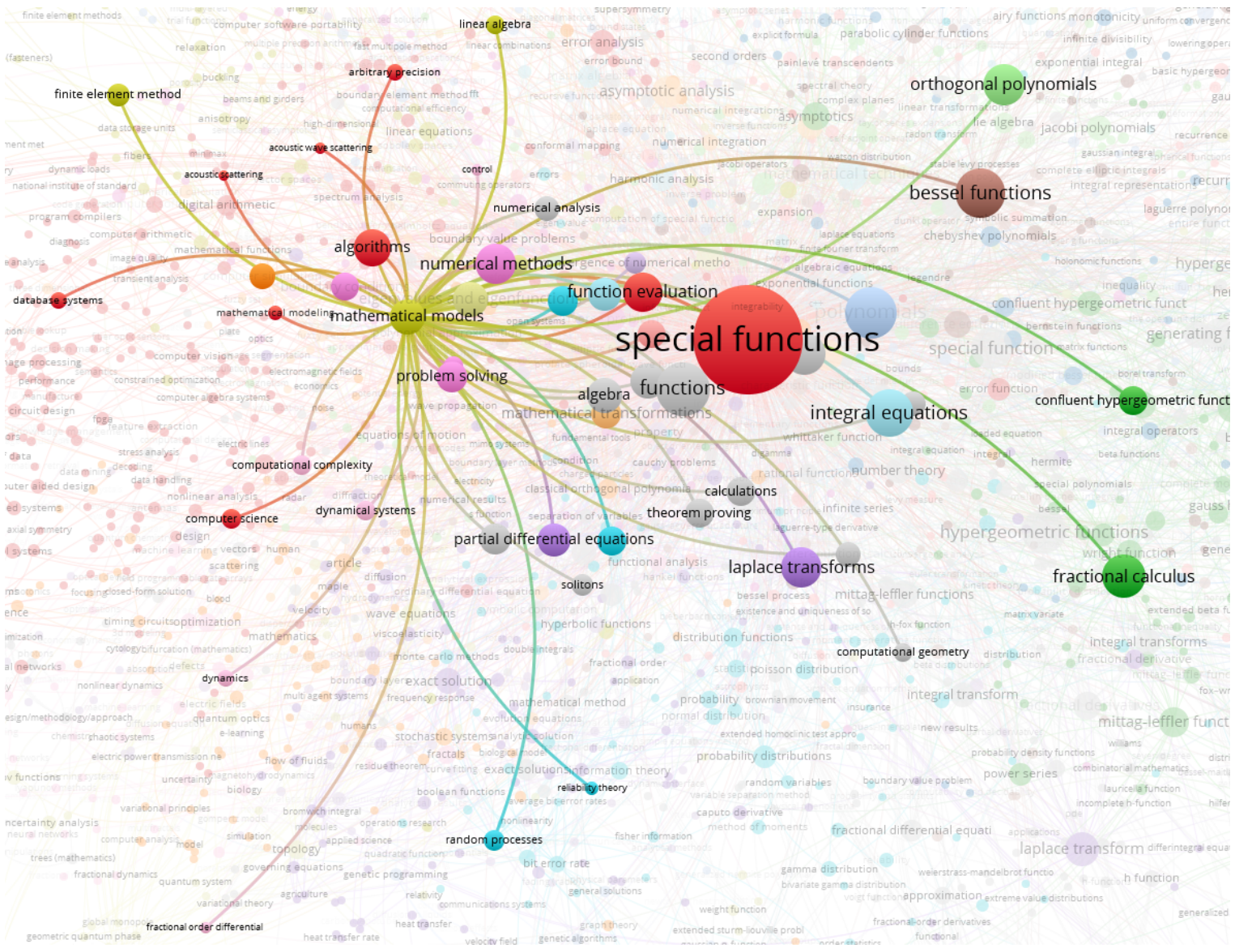

Figure 16.: The links of "mathematical models" in the studied keyword network.

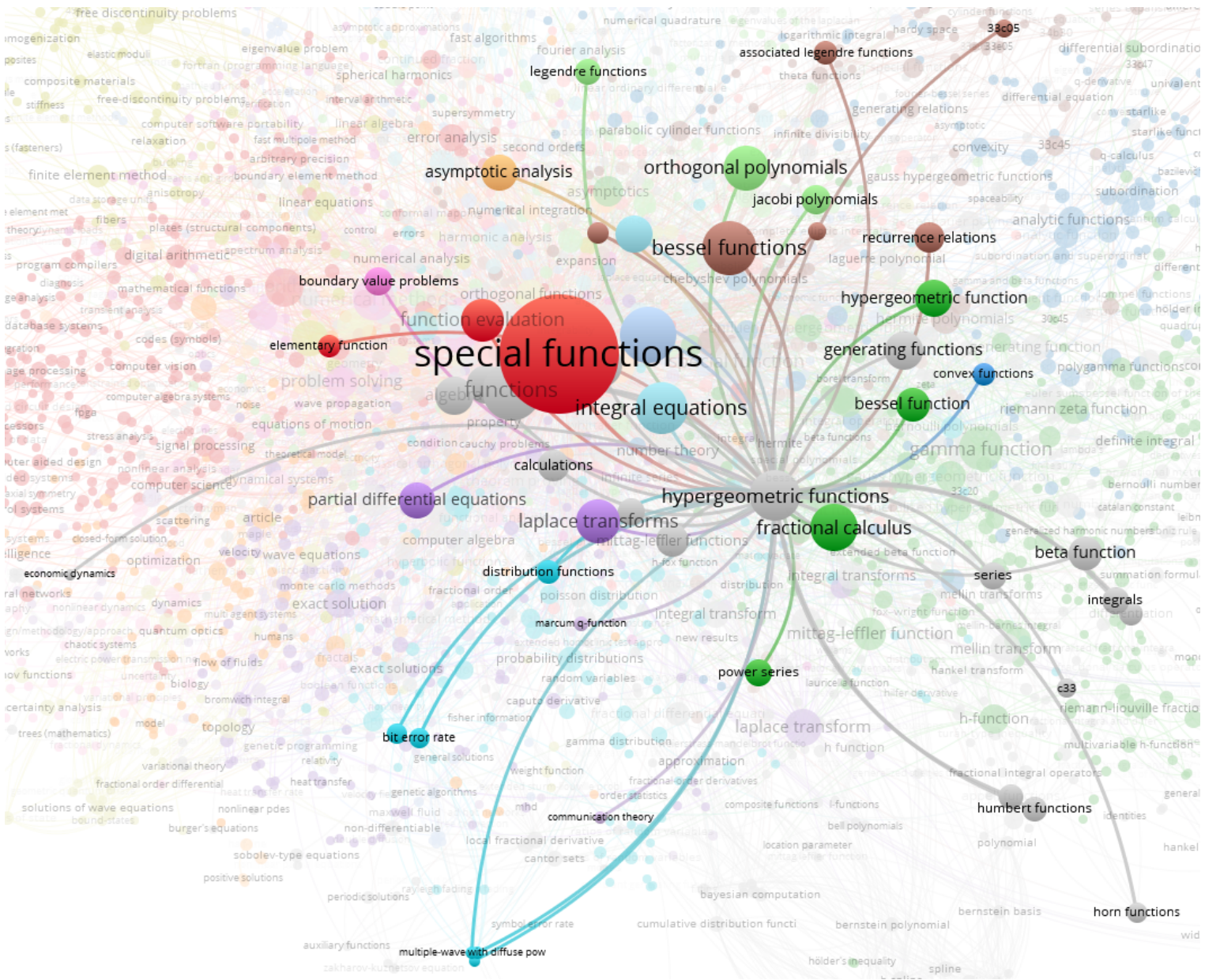

Figure 17.: The links of "hypergeometric function" in the studied keyword network.

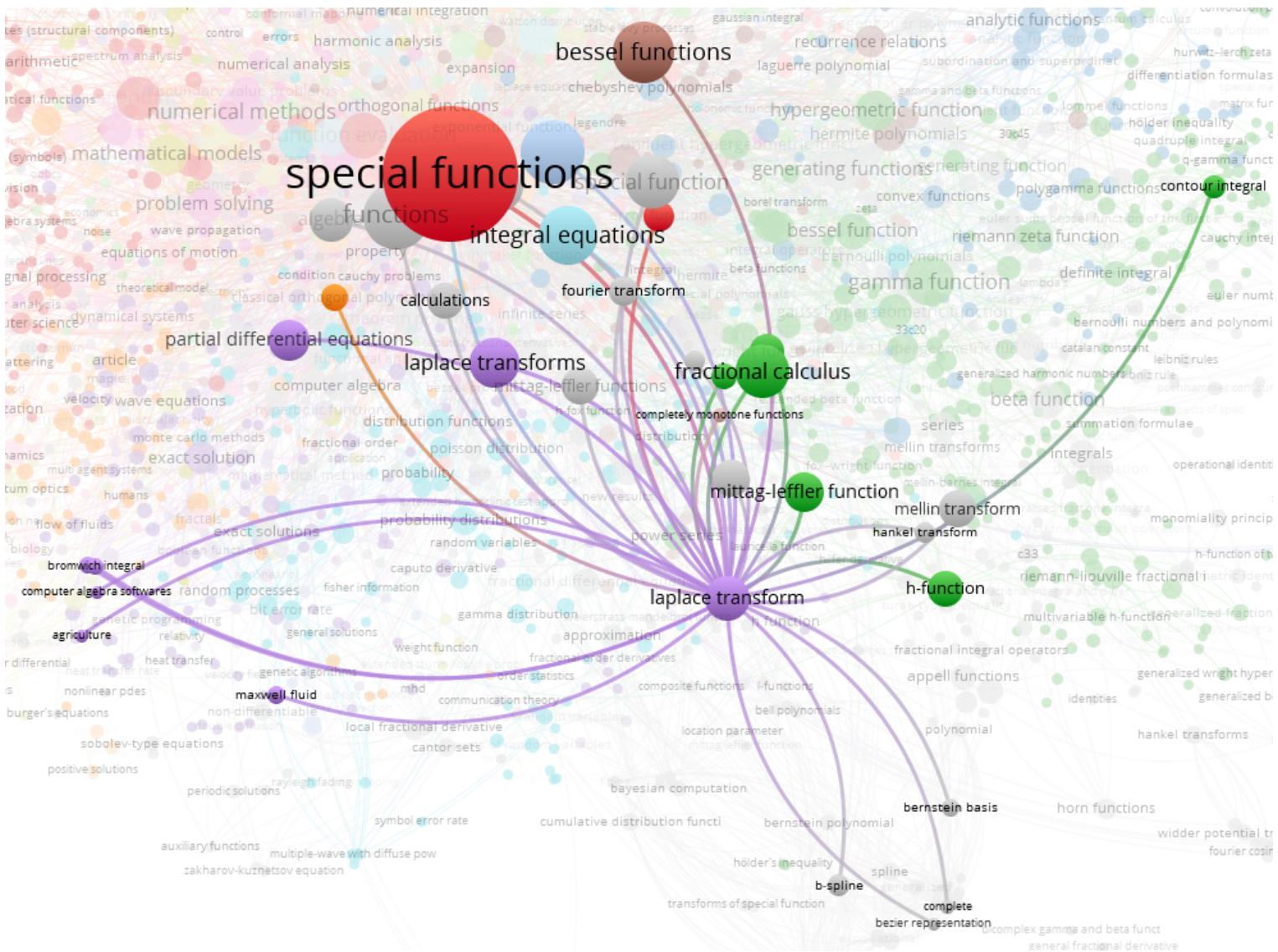

Figure 18.: The links of "Laplace transform" in the studied keyword network.

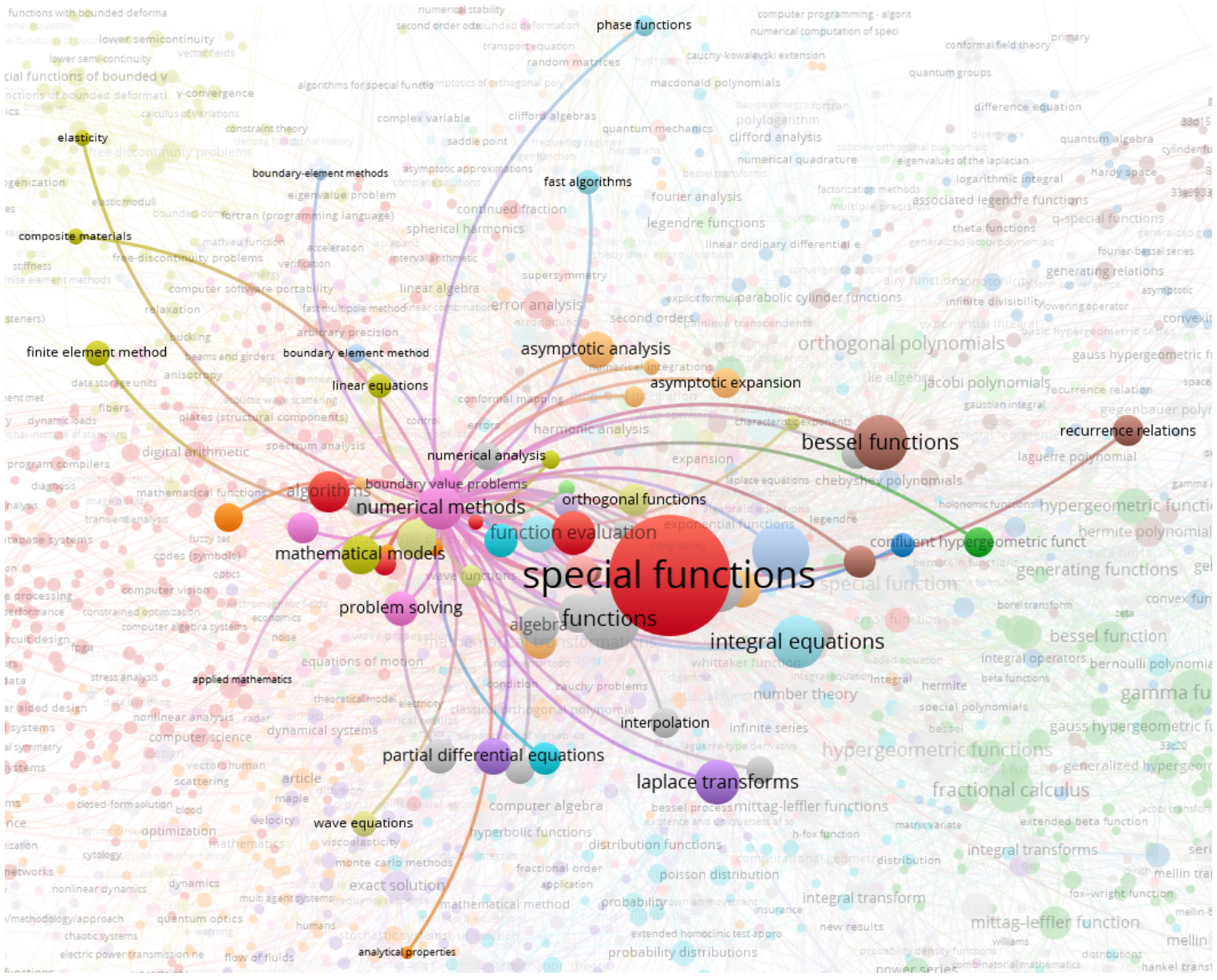

Figure 19.: The links of "numerical methods" in the studied keyword network.

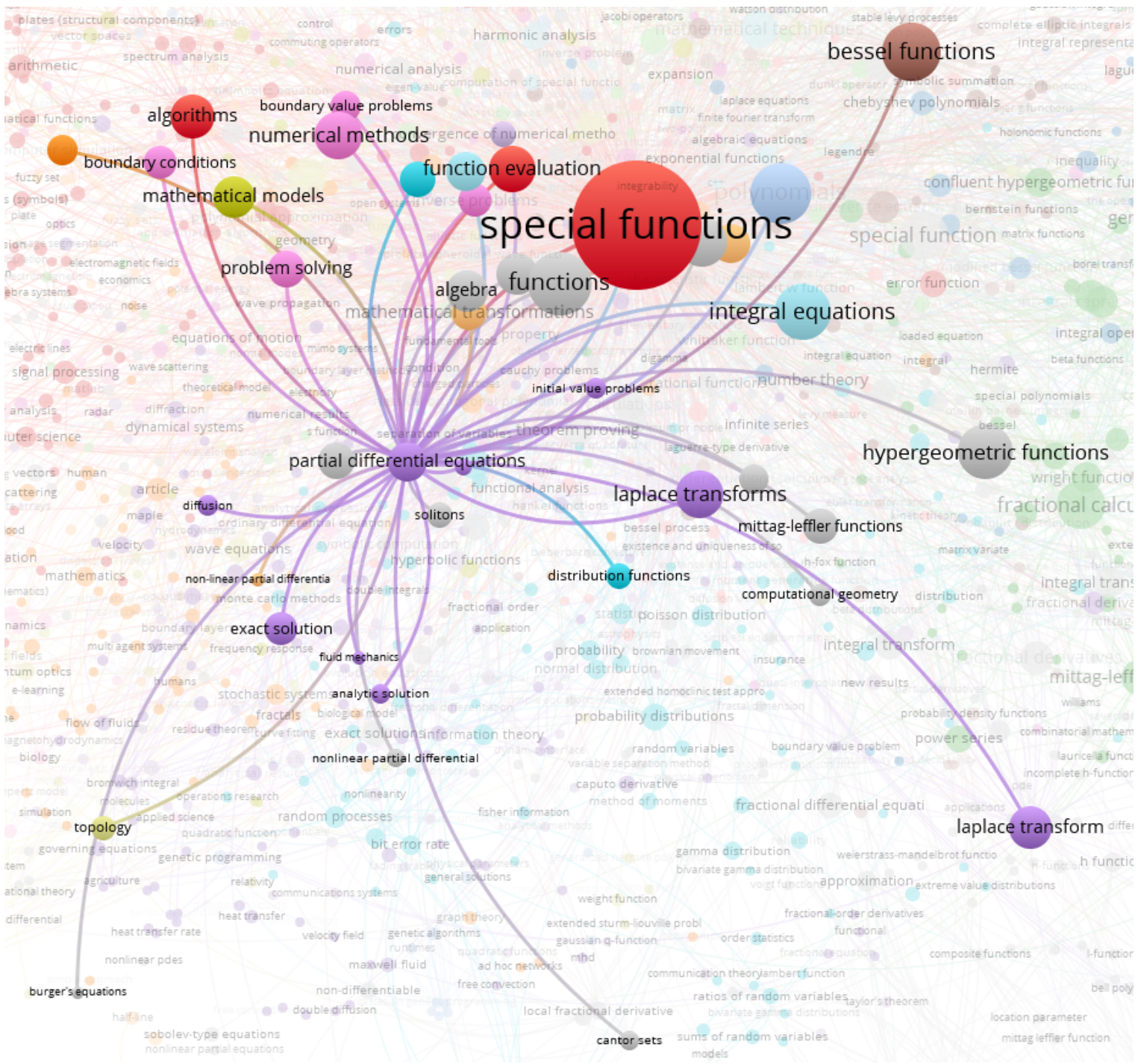

Figure 20.: The links of "partial differential equations" in the studied keyword network.

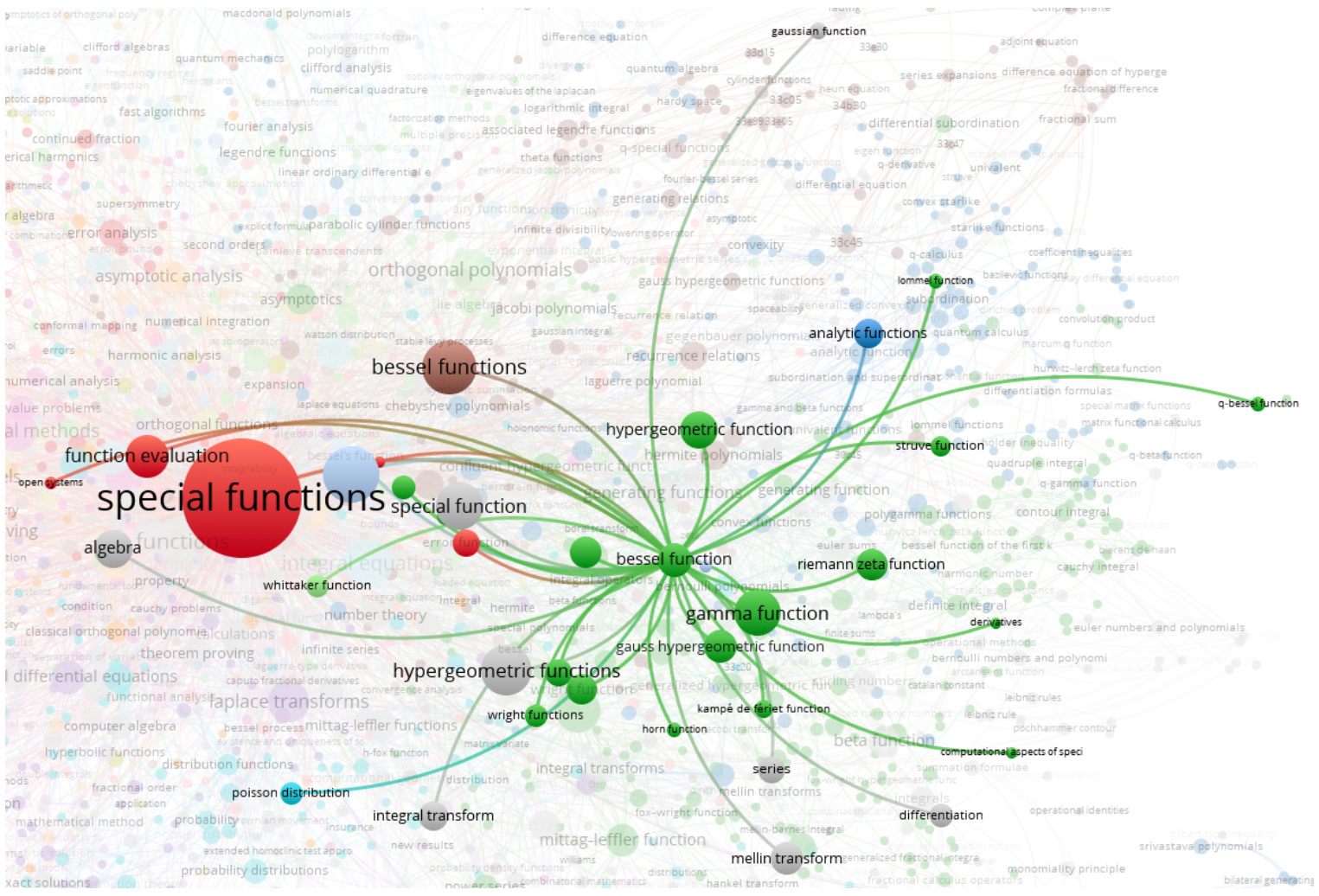

Figure 21.: The links of "Bessel function" in the studied keyword network.

## 4. Geometric modeling and special functions

### *4.1. Unpopularity of special functions in geometric modelling*

Unfortunately, special functions are not popular in geometric modelling [4], a branch of applied mathematics and computational geometry that studies methods and algorithms for the mathematical description of shapes.

One of the first approaches to use general families of special functions was done in a few works. Ziatdinov et al. found an analytical representation of log-aesthetic curves in terms of incomplete gamma functions in [5]. The superspirals proposed in [6] use the Gaussian hypergeometric function for the curvature representation, and the curve segments are computed numerically by the Gauss-Kronrod numerical integration method. Inoguchi et al. studied superspirals derived from the Kummer confluent hypergeometric function in [7,8].

The node 'Geometric modelling' appears to be a relatively minor topic within the studied network, with only six occurrences, indicating that it does not occur frequently in the dataset (see Figs. 22 and 23). Although it has 36 links connecting it to other nodes, this is below the average connectivity of the network. Its total link strength of 106 is relatively low, suggesting that although it has some links, they may not be as intense or frequent as other keywords. This low number of occurrences may indicate that the special functions are not commonly used in geometric modelling. However, its connections to 36 nodes may be worth exploring to understand its role in the broader research landscape and its relationships with more prominent topics.

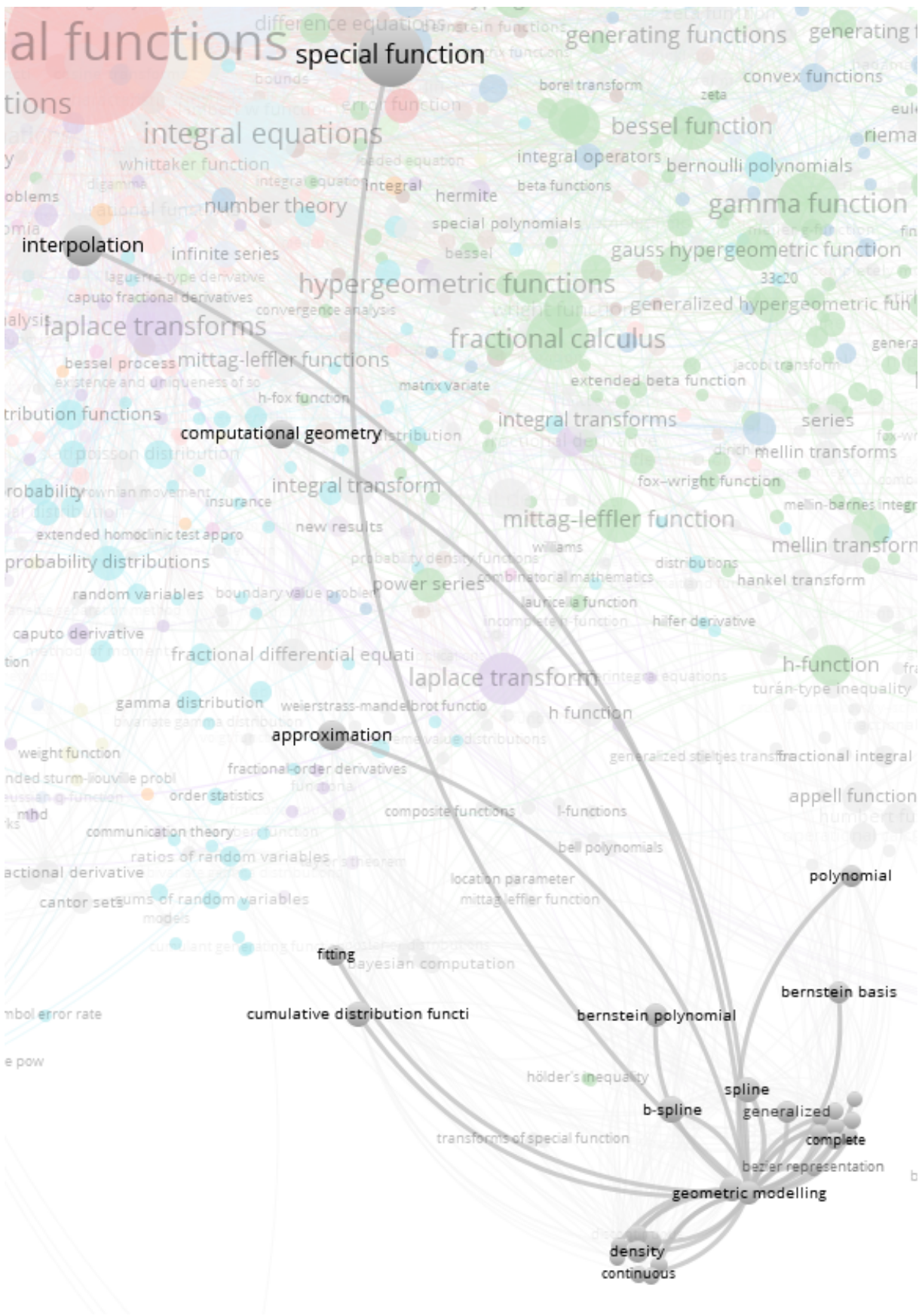

Figure 22.: The links of "geometric modelling" in the studied keyword network.

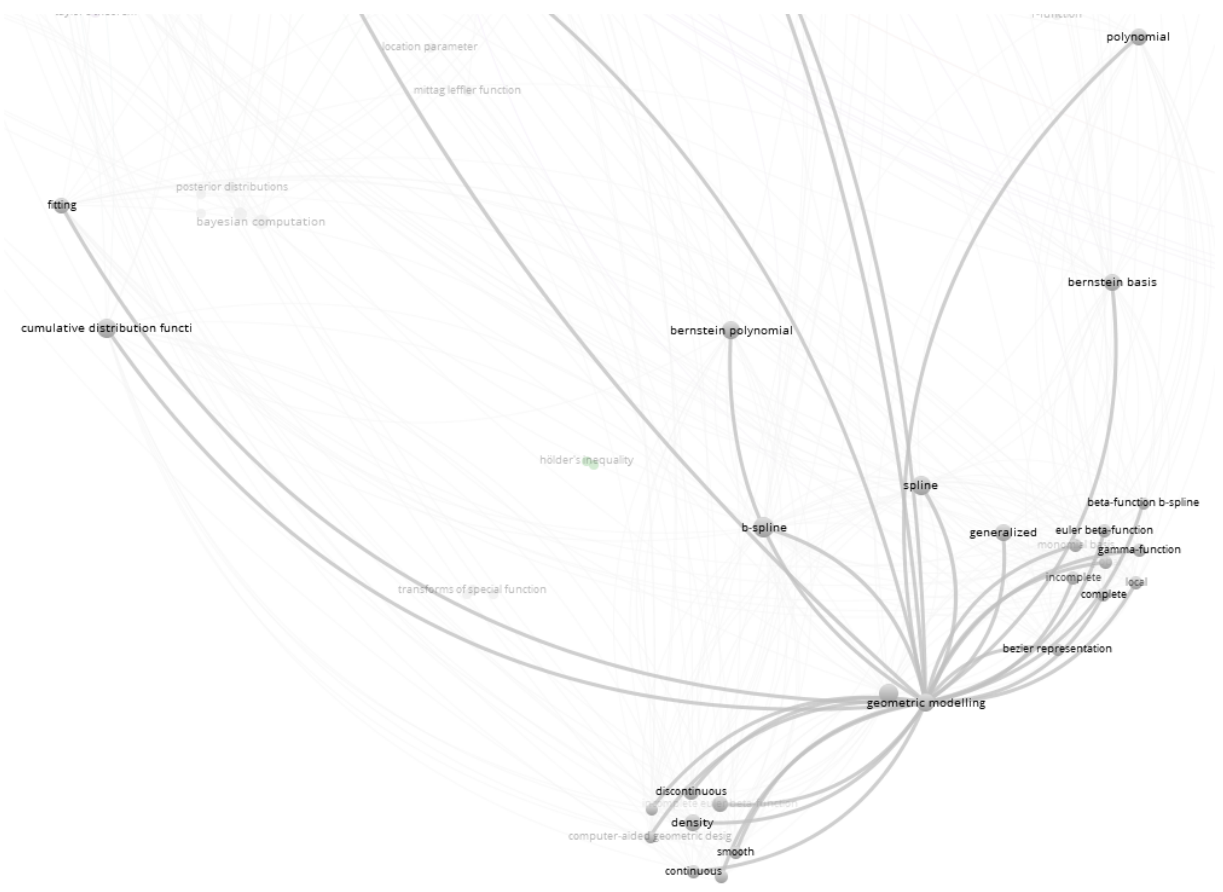

Figure 23.: The links of "geometric modelling" in the studied keyword network.

### *4.2. Special functions in Julia programming language*

Special functions in the Julia programming language can be computed using the following packages [9]:

- **SpecialFunctions.jl:** Special functions like `airyai(x)`, `airyaiprime(x)`, `airyaix(x)`, `airyaiprimex(x)`, `airybix(x)`, `airybiprimex(x)`, `airybi(x)`, `airybiprime(x)`, `besselh(nu,k,x)`, `besselix(nu,x)`, `besseli(nu,x)`, `besselj(nu,x)`, `besselj0(x)`, `besselj1(x)`, `besseljx(nu,x)`, `besselk(nu,x)`, `besselkx(nu,x)`, `bessely(nu,x)`, `bessely0(x)`, `bessely1(x)`, `besselyx(nu,x)`, `dawson(x)`, `digamma(x)`, `erf(x)`, `erfc(x)`, `erfcinv(x)`, `erfcx(x)`, `erfi(x)`, `erfinv(x)`, `eta(x)`, `hankelh1(nu,x)`, `hankelh1x(nu,x)`, `hankelh2(nu,x)`, `hankelh2x(nu,x)`, `zeta(x)`.
- **Bessels.jl:** Bessel functions for real arguments and orders that include `airyai(x)`, `airyaiprime(x)`, `airybi(x)`, `airybiprime(x)`, `besselh(nu, k, x)`, `besseli(nu, x)`, `besseli0(x)`, `besseli1(x)`, `besselj(nu, x)`, `besselj0(x)`, `besselj1(x)`, `besselk(nu, x)`, `besselk0(x)`, `besselk1(x)`, `bessely(nu, x)`, `bessely0(x)`, `bessely1(x)`, `Bessels.gamma(x)`, `Bessels.sphericalbesseli(nu, x)`, `Bessels.sphericalbesselk(nu, x)`, `hankelh1(nu, x)`, `hankelh2(nu, x)`, `sphericalbesselj(nu, x)`, `sphericalbessely(nu, x)`.
- **LogExpFunctions.jl:** A package for various special functions based on logarithmic and exponential functions.
- **HypergeometricFunctions.jl:** A package for the calculation of the generalised hypergeometric functions.
- **Struve.jl:** Struve functions such as `struveh(nu, x)`, `struvek(nu, x)`, `sruvel(nu, x)`, `struvem(nu, x)`, `Struve.H0_fast(x)`, `Struve.H1_fast(x)`.
- **LambertW.jl:** Lambert W (also known as the omega function or product logarithm) mathematical function, such as `lambertw(x)`, `lambertw(x,k)`, `LambertW.finv(lambertw)`, `lambertwbp(x,k)`, `omega`.

which can be installed using the command `import Pkg; Pkg.add("Package`

name"). Not all the special functions popular in Scopus-indexed documents are found here, but some can be found on GitHub at https://github.com/JuliaMath.

Figures 24, 25 show special function, their curvature, and the curvature of curvature functions. When a designer analyses shapes, the curvature function provides valuable insight into the geometric properties of a shape, helping to determine how a curve or surface bends at different points, identifying areas of high or low curvature, inflexion or singular points, and symmetry regions. This information is helpful in optimising designs for aesthetics, functionality and manufacturability.

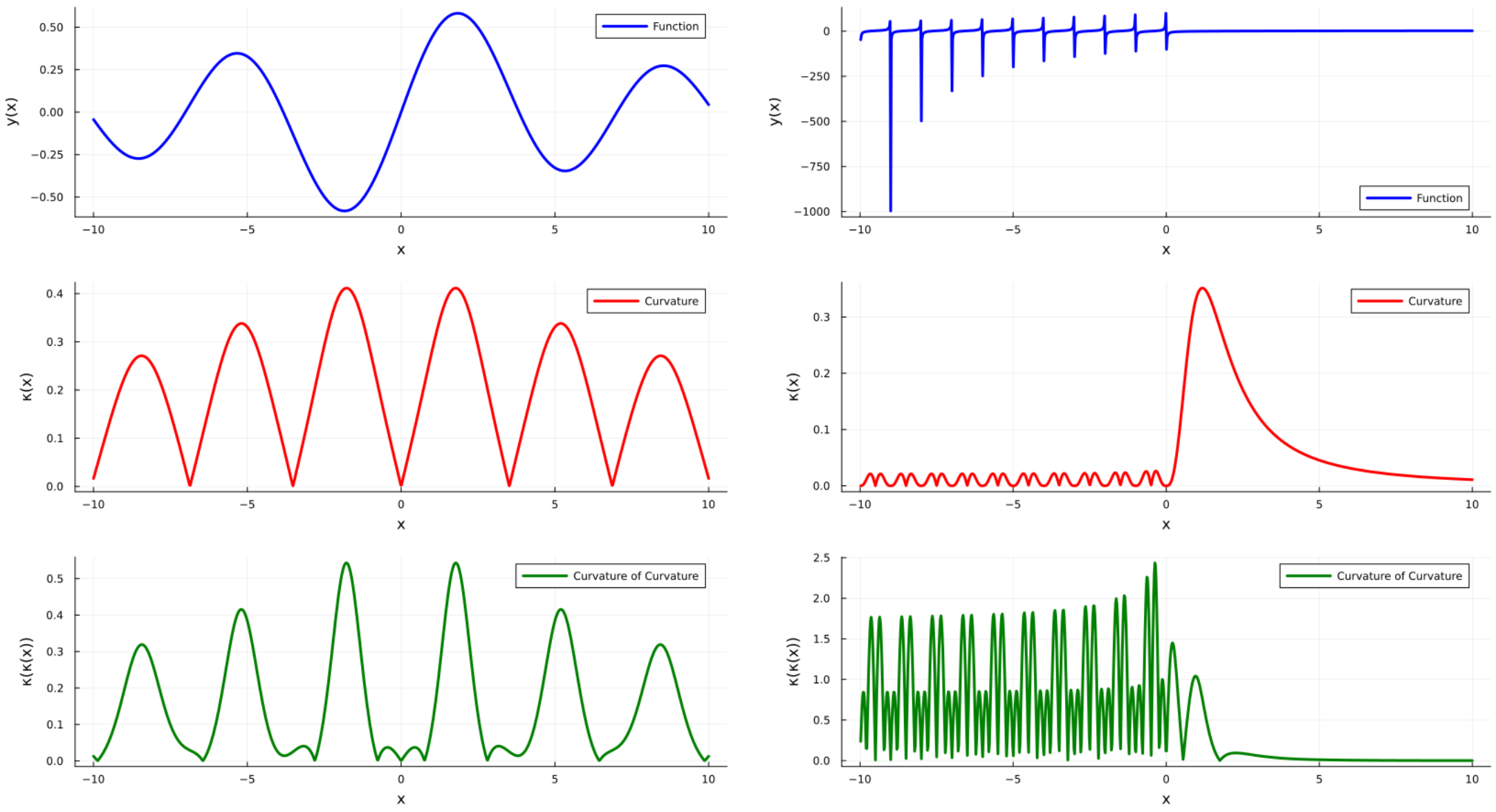

((a)) Bessel function of the first kind of order $\nu = 1$ at $x$.

((b)) Digamma function.

Figure 24.: Special functions, their curvature functions $\kappa(x)$ and the curvature of curvature functions $\kappa(\kappa(x))$.

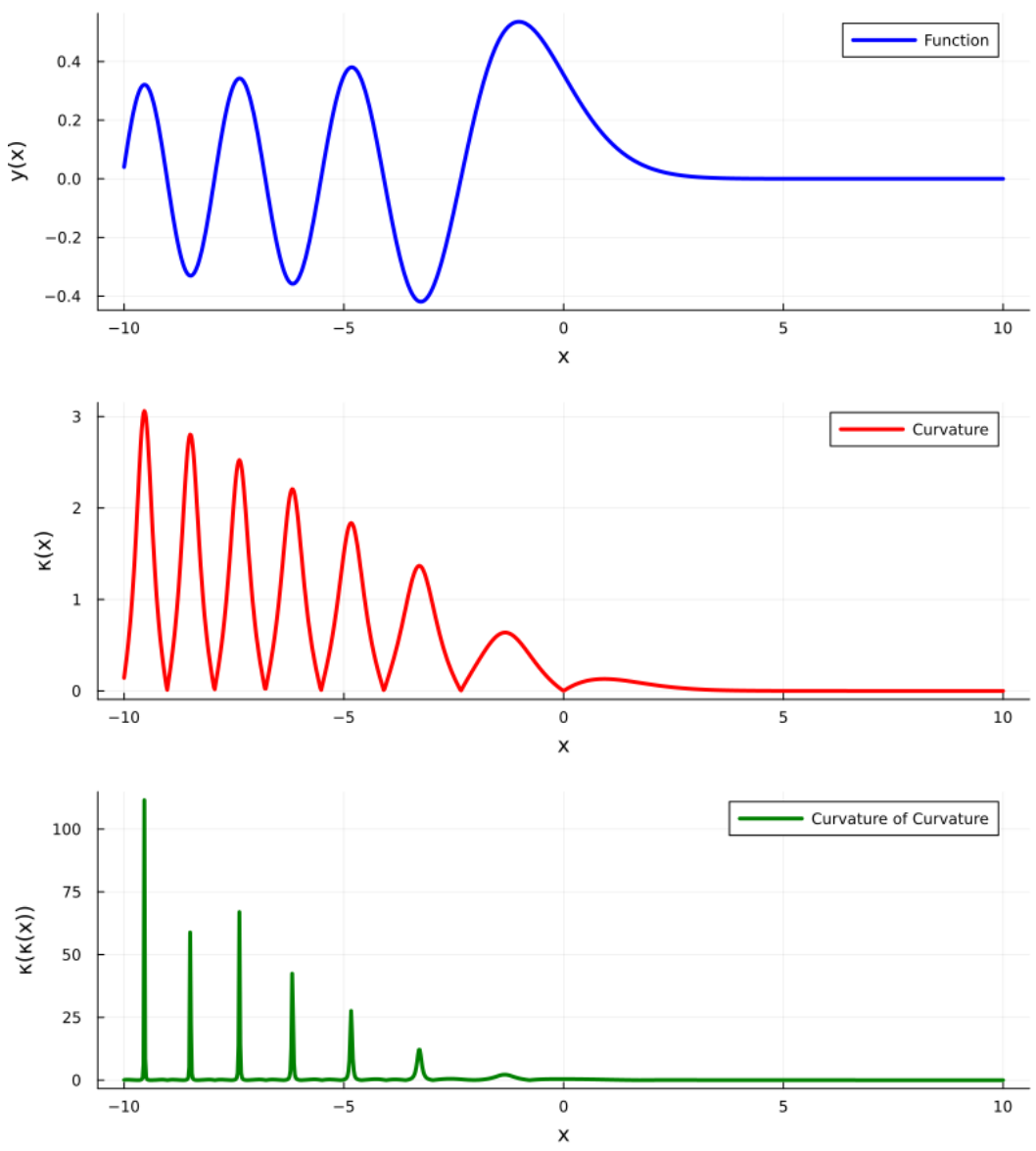

((a)) Airy function.

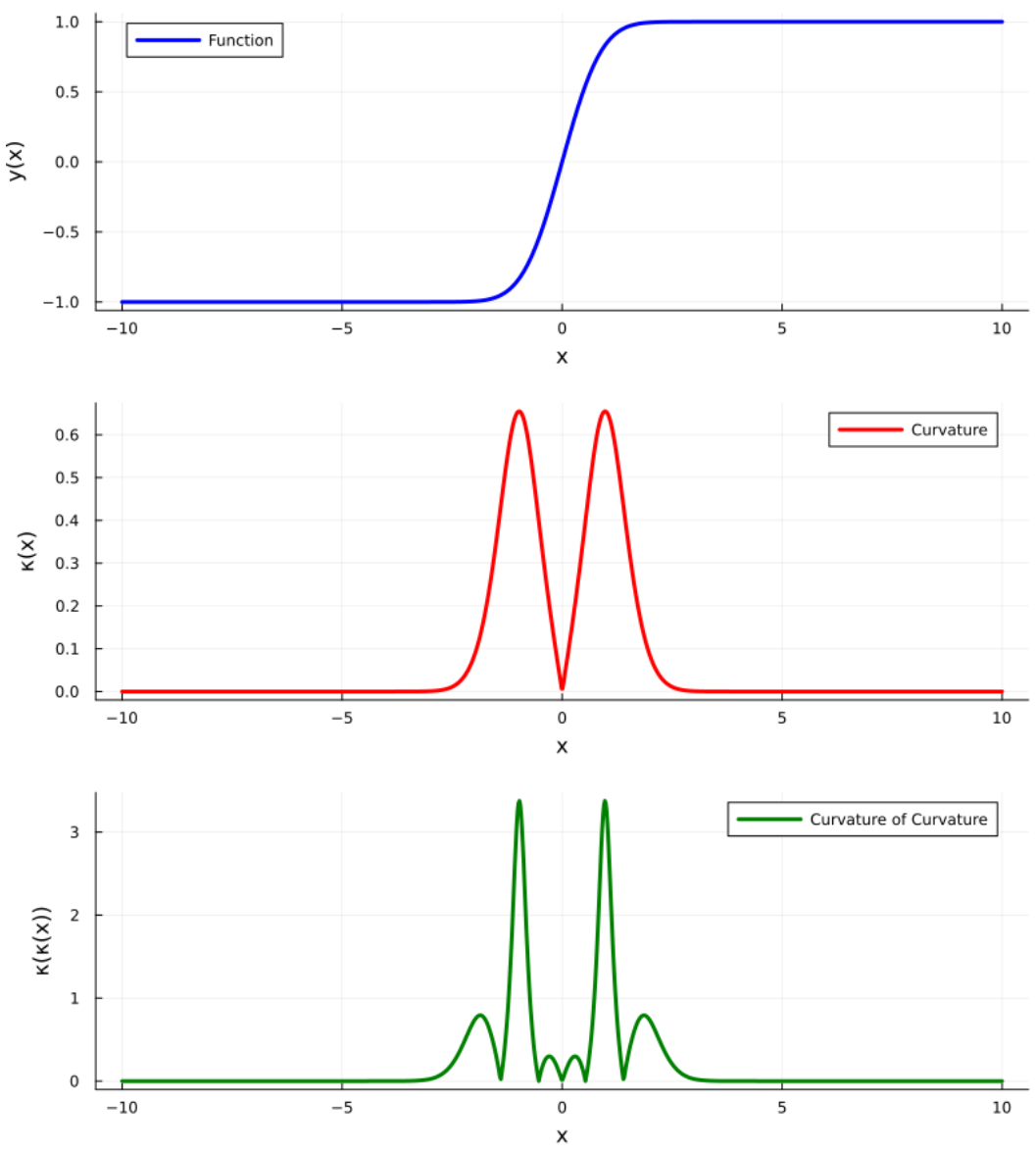

((b)) Error function.

Figure 25.: Special functions, their curvature functions $\kappa(x)$ and the curvature of curvature functions $\kappa(\kappa(x))$.

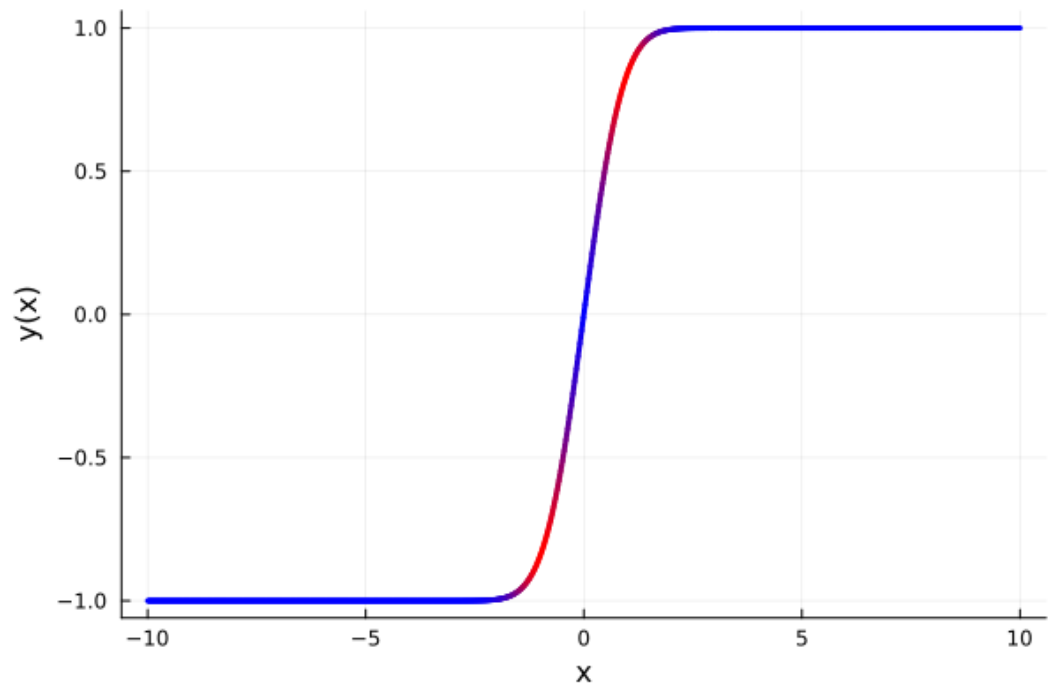

((a)) Error function.

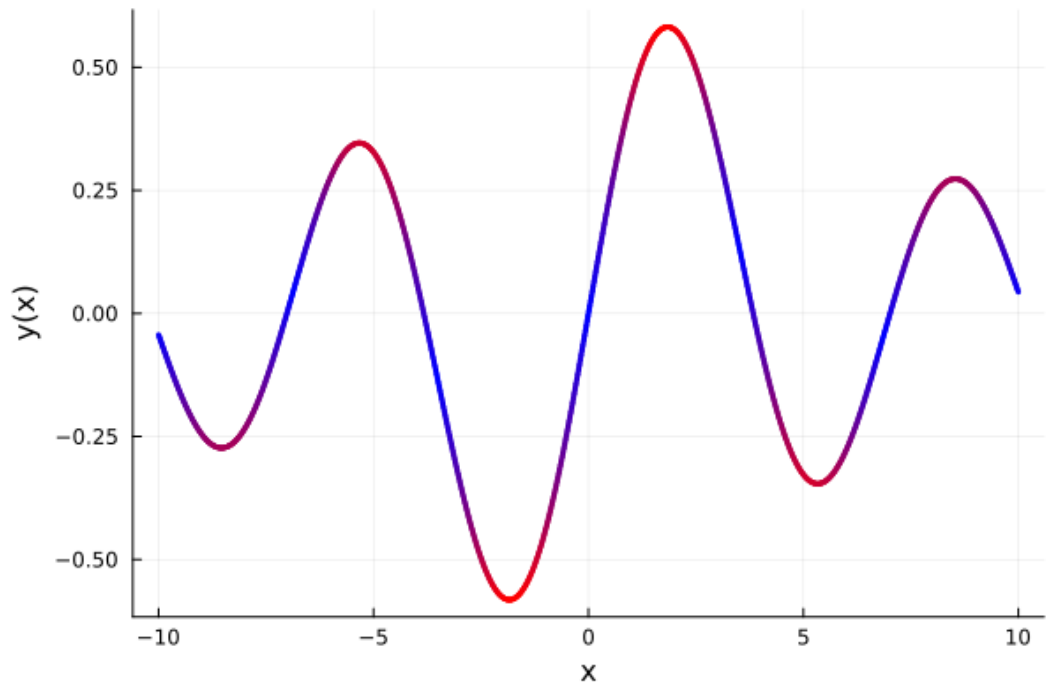

((b)) Bessel function `besselj(1, x)`.

Figure 26.: Visualisation of special functions where red colour shows high absolute curvature values and blue colour shows low absolute curvature values at relative points.

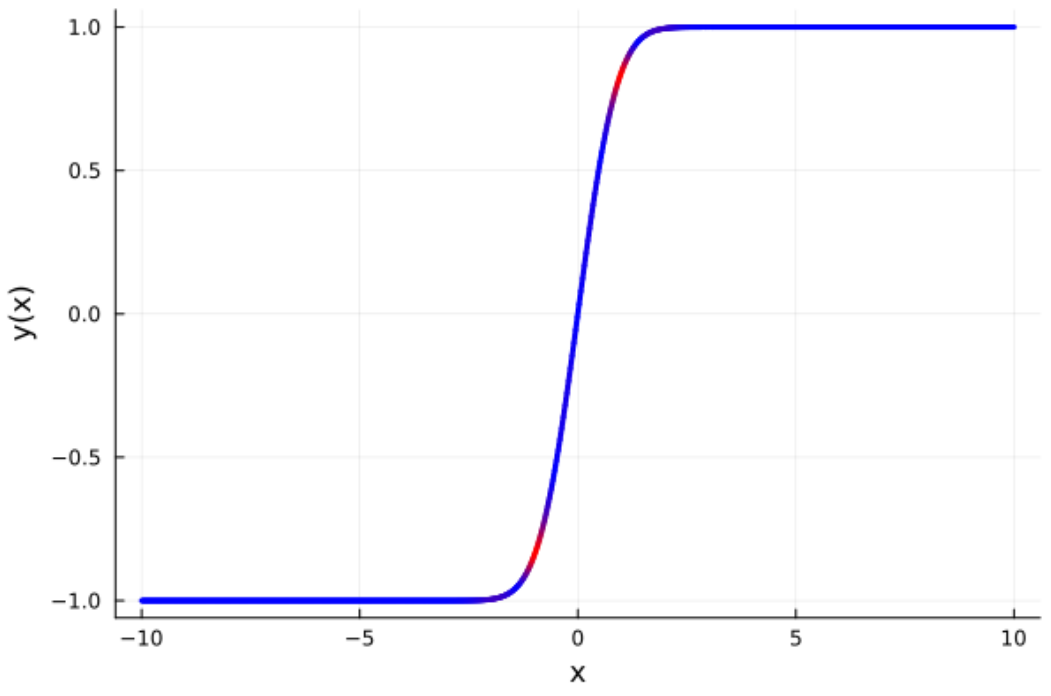

((a)) Error function.

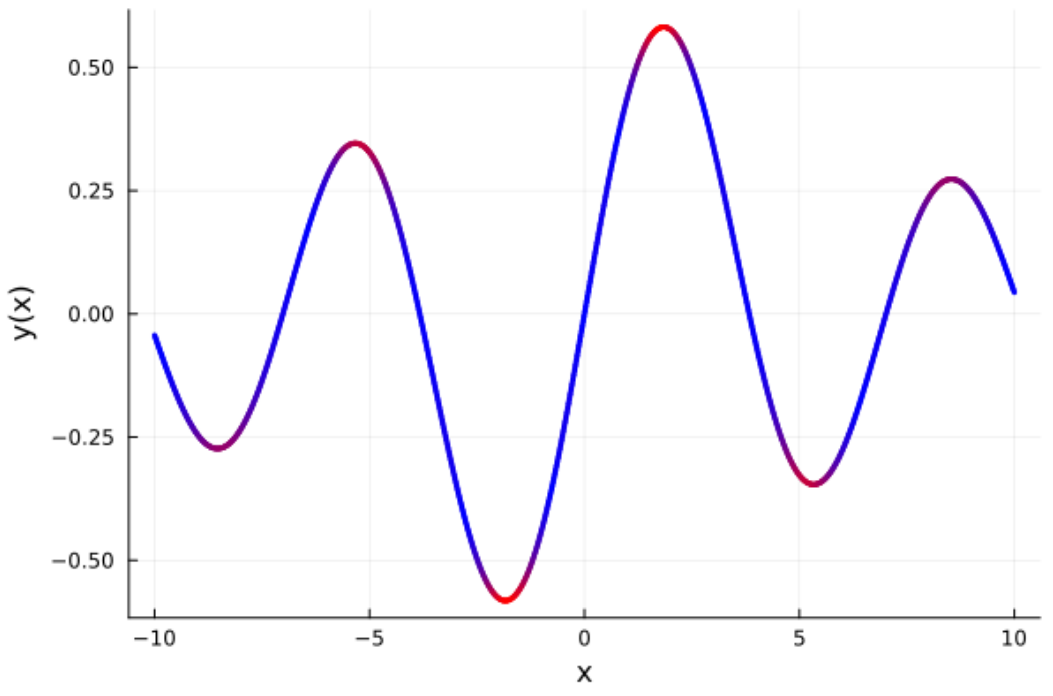

((b)) Bessel function `besselj(1, x)`.

Figure 27.: Visualisation of special functions where red colour shows high absolute values of curvature of curvature function and blue colour shows low absolute values of curvature of curvature function at relative points.

## 5. Conclusions

This work aimed to create different network visualisations related to ”special function” keyword-related document data extracted from the Scopus database. By visually analysing these visualisations, researchers can see strong connections or lack of connections between different topics related to special functions, also they can help to find appropriate journals, co-authors, supervisors and applications for research on special functions. We also discussed various special functions of the Julia programming language that can be helpful to researchers in special function theory and its applications, for example, in the field of geometric modelling.